# Comparing Active Anomaly Detection Frameworks: From Light Curves to Images

Sreevarsha Sreejith[a], Robert C. Nichol[a]

[a]*School of Maths and Physics, University of Surrey, Stag Hill, University Campus, Guildford, GU2 7XH, UK*

**Abstract**

The increasing volume and complexity of data produced by modern astronomical surveys are making comprehensive visual inspection increasingly impractical, thus motivating the development of automated methods for identifying rare and unusual objects. Active anomaly detection (A-AD) combines unsupervised anomaly detection with iterative user feedback, allowing searches to adapt to a user's scientific interests while retaining sensitivity to unexpected objects. In this work, we systematically compare three A-AD frameworks, Astronomaly, Protege, and PineForest, across two astronomical data modalities: Zwicky Transient Facility light curves and Euclid galaxy images. We evaluate their behaviour under experimental conditions, including initial anomaly recovery and the effect and transferability of prior user labels between methods. The three frameworks exhibit substantially different behaviour across the two datasets: in the baseline experiments, Protege, PineForest, and Astronomaly recovered 83, 78, and 38 variable stars, respectively, within the first 100 light-curve objects inspected, while PineForest recovered 70 merging or morphologically disturbed galaxies within the first 100 Euclid images, compared with 29 and 19 for Astronomaly and Protege. Experiments using prior labels further show that their effectiveness depends not only on the availability of prior information, but also on the composition of the prior sample, with their impact varying between the A-AD methods. Finally, we demonstrate how the complementary behaviour of Protege and PineForest can be combined in a hybrid search of 10,000 Euclid images, recovering 799 morphologically unusual objects. These include 29 retained strong lens systems and candidates, of which 9 candidates have no counterparts in the published Euclid Q1 strong lens catalogues searched, and 202 visually classified ring galaxies. These results highlight the importance of selecting A-AD strategies according to both the scientific objective and the properties of the dataset, and demonstrate their potential for efficient, discovery-driven exploration of forthcoming astronomical surveys.

## 1. Introduction

Astronomical discoveries have always followed the path of human observation and visual examination, from Democritus' early attempts to explain the nature of the Milky Way in ancient Greece to Hubble's observations at Mount Wilson Observatory in the early $20^{th}$ century. This tradition continues with today's state-of-the-art space- and ground-based observatories equipped with increasingly sophisticated instrumentation. The prospect of spotting and recognising something new through direct inspection of astronomical observations has remained one of the defining aspects of observational astronomy. Serendipitous discoveries have also played an important role in advancing astronomy, with unexpected objects and phenomena often identified in observations obtained for entirely different scientific purposes, notable examples including the discovery of pulsars (Bell Burnell, 2007) and Hanny's Voorwerp (Lintott et al., 2009). Preserving the ability to recognise such unexpected discoveries is therefore an important consideration as astronomical datasets continue to grow.

However, that era in which astronomical datasets could be visually examined might be coming to an end in astronomy, with the advent of new telescopes and surveys, both current and upcoming, such as DES (Dark Energy Survey, Abbott et al. 2018), JWST (James Webb Space Telescope, Gardner et al. 2006), Euclid (Refregier et al., 2010), Nancy Grace Roman Space Telescope (previously WFIRST, Green et al. 2012), Vera C. Rubin Observatory Legacy Survey of Space and Time, (LSST Science Collaboration et al. 2009; Ivezić et al. 2019) which are and which will compile unprecedented amounts of data that will render comprehensive visual inspection impossible. Citizen-science initiatives such as Galaxy Zoo (Lintott et al., 2008) have substantially extended the scale at which visual inspection can be performed by distributing classification tasks among large numbers of volunteers. Nevertheless, the rapidly increasing volume of survey data presents a challenge even for such distributed approaches.

This increase in data volume affects several aspects of astronomical analysis such as in detection of objects, their classification, image analysis, recommendation of objects for follow-up spectroscopy etc. Due to this big data overload, using automated statistical and machine learning methods of data analysis is becoming commonplace in astronomy, especially using machine learning methods to detect and classify objects, to deblend and denoise sources etc. A more recent and tantalising application of machine learning methods to astronomical data is using outlier detection algorithms to search these vast data sets for both 'known' and 'unknown' unknowns, more commonly referred to as anomalies. Such anomalies may correspond to rare but known objects like superluminous supernovae (Gal-Yam, 2012), well known and documented objects that can help us constrain cosmological parameters like Type Ia Supernovae (Filippenko, 1997), or something entirely new, exotic and unforeseen. *Anomaly detection* (AD) therefore offers a means of preserving the potential for serendipitous discovery in datasets that can no longer be comprehensively inspected by human eye.

AD algorithms are generally unsupervised machine learning algorithms, which means that they do not require the input data to be labelled (in contrast with supervised algorithms which are the most widely used machine learning algorithms that require a labelled training set on which to train the data). There are several such algorithms that are used for outlier or novelty detection in data sets, such as Isolation Forest (Liu et al., 2008), Local Outlier Factor (Breunig et al., 2000), Gaussian Mixture Model (McLachlan, 2000) and One-class Support Vector Machines (Schölkopf et al., 1999).

These methods or variations thereof have been used in astronomy to detect and identify anomalies in galaxy spectra (Baron and Poznanski 2017; Škoda et al. 2020), transients (Martínez-Galarza et al. 2021; Pruzhinskaya et al. 2019; Zhang and Zou 2018; Gupta et al. 2024; Muthukrishna et al. 2022; Chaini et al. 2025), photometric redshift estimation (Hoyle et al. 2015), variable stars (Rebbapragada et al. 2009; Nun et al. 2014; Giles and Walkowicz 2019; Malanchev et al. 2021) and images (Hocking et al. 2018; Polsterer et al. 2019; Doorenbos et al. 2021; Margalef-Bentabol et al. 2020) among several others. More recently, self-supervised representation learning has been combined with classical AD algorithms, enabling complex astronomical images to be embedded into compact high-dimensional feature spaces before AD is performed, thereby substantially improving the effectiveness of image-based discovery pipelines (Hayat et al. 2021; Bardes et al. 2022; Mohale and Lochner 2024).

While AD by itself has shown considerable promise for analysing astronomical data, the highest ranked anomalies are not always the ones that are most scientifically interesting. In practice, AD algorithms identify a mixture of unusual astrophysical objects, data artefacts and other statistical outliers, making it difficult to distinguish genuine anomalies from bogus detections. One way to get around this is by designing bespoke algorithms that target a particular object class, however, the definition of 'interesting' varies depending on the scientific objective (Pruzhinskaya et al., 2025). Furthermore, such approaches are generally more useful when searching for *known unknowns*. When the goal is to discover new classes of objects or *unknown unknowns*, the features of the target class are unavailable a priori.

This is where *Active Learning* (AL, Settles 2012) becomes particularly relevant. AL incorporates feedback from a human expert by iteratively updating the learning model using a small number of labelled examples. Thus the search can be progressively guided towards scientifically relevant objects while remaining flexible enough to identify the unexpected. AL has been used successfully in a variety of fields such as natural language processing (Thompson et al., 1999) and spam classification (DeBarr and Wechsler, 2009). In astronomy, it has been used in applications such as supernova photometric classification and optimisation of spectroscopic follow-up (Gupta et al. 2016; Ishida et al. 2019; Kennamer et al. 2020; Möller et al. 2025), photometric redshift estimation (Vilalta et al., 2017), stellar population parameter determination (Solorio et al., 2005), variable star classification (Richards et al., 2012), galaxy morphology classification (Walmsley et al., 2020) and telescope choice optimisation (Xia et al., 2016) among others.

In recent years the combination of AL and AD has become increasingly popular in astronomy, termed *Active Anomaly Detection* (A-AD henceforth). Instead of relying solely on an unsupervised ranking like AD, A-AD methods iteratively integrate user feedback to guide the model towards objects of scientific interest based on user-defined criteria. Several implementations of this paradigm have recently been developed for astronomical applications, including Astronomaly (Lochner and Bassett, 2021), Coniferest (Kornilov et al., 2025) and AnomalyMatch (Gómez et al., 2026), demonstrating its effectiveness for interactive exploration of large survey datasets. These frameworks have subsequently been applied to a wide range of astronomical discovery problems, including large scale galaxy searches (Etsebeth et al., 2024), time-domain anomaly discovery (Volnova et al. 2024; Pruzhinskaya et al. 2023), anomaly detection on Hubble archive images (O'Ryan and Gómez, 2025) and the identification of image artefacts in ZTF images (Sreejith et al., 2026).

Despite these promising developments, the relative strengths and weaknesses of different A-AD frameworks remain poorly understood. In particular, there has been no systematic comparison of existing methods under identical experimental conditions or across different astronomical data modalities. A-AD methods differ in how they represent their inputs: some operate on pre-extracted feature vectors, whereas others (e.g. AnomalyMatch) learn directly from images. This comparison is intentionally restricted to frameworks that operate on pre-extracted feature representations. End-to-end approaches, such as AnomalyMatch, jointly learn feature representations and active learning strategies, making direct comparisons difficult because differences in performance may arise from either the representation learning or the active anomaly detection algorithm itself. By fixing the feature representation across methods within each dataset, the present study isolates the effect of the active anomaly detection strategy.

Separating feature extraction and A-AD enables the usage of advanced representation learning to be applied independently of the A-AD algorithm itself. Once an appropriate feature representation has been extracted, these methods are largely agnostic to the underlying data modality, which allows their behaviour to be compared across light-curve and image datasets within the same experimental conditions. Understanding how different A-AD strategies behave across diverse datasets is essential for selecting appropriate methods for future large-scale astronomical surveys.

In this work, we present a systematic comparison of three A-AD frameworks: PineForest, Astronomaly and Protege. Using both time-domain light curve data and higher-dimensional image feature embeddings, we investigate their anomaly recovery efficiency, the transferability of prior information between methods and their behaviour across different astronomical data representations. Finally, we demonstrate the practical application of A-AD by combining the complementary strengths of PineForest and Protege to explore morphologically unusual objects from Euclid imaging, illustrating the potential of hybrid A-AD strategies for exploration in the era of large-scale astronomical surveys.

## 2. Active Anomaly Detection Frameworks

The three A-AD frameworks compared in this work are Astronomaly (Lochner and Bassett, 2021), its extension Protege (Lochner and Rudnick, 2025), and PineForest (Kornilov et al. 2025; Ishida et al., in preparation). Their respective A-AD strategies and implementations used in this work are described below.

### *2.1.* PineForest

PineForest is an A-AD method developed by the SNAD collaboration that builds upon the Isolation Forest (Liu et al., 2008) framework while retaining the underlying decision tree structure. Rather than modifying anomaly scores directly, PineForest adopts a 'tree-filtering' strategy in which individual trees are evaluated according to their ability to distinguish labelled anomalous objects from labelled regular objects. The usefulness of individual trees is estimated using the path lengths of labelled objects through each tree, rewarding trees that isolate known anomalies more effectively than known regular objects. Trees with the lowest usefulness scores are removed and replaced by newly generated random trees, maintaining a constant number of trees in the forest.

During an interactive A-AD session, the user is iteratively presented with the highest ranked candidate anomaly in each iteration and assigns each object a label of either 'ANOMALY' or 'REGULAR'[1]. After each new label is obtained, the scores of all trees are recomputed using the complete set of accumulated labels. The lowest scoring trees are discarded and replaced, and anomaly scores are recalculated for the dataset. Through this iterative process, the forest gradually adapts to the user's definition of anomalous behaviour, while retaining computational efficiency as the number of accumulated labels increases.

PineForest is available as part of the `coniferest`[2] package alongside a reimplementation of `scikit-learn`'s Isolation Forest and the A-AD algorithm Active Anomaly Discovery (Das et al. 2017; Ishida et al. 2021). [3]

### *2.2.* Astronomaly

Astronomaly is a general-purpose framework for A-AD in astronomical datasets that combines unsupervised anomaly detection with human-in-the-loop active learning (Lochner and Bassett, 2021). The framework can operate on a variety of astronomical data, such as images, spectra, and light curves, and provides a modular architecture for feature extraction, anomaly detection, visualisation, and user feedback. It consists of a Python backend for data processing and machine learning together with a JavaScript-based web interface through which users can inspect objects, visualise feature spaces, and provide feedback.

The anomaly detection stage is performed independently of the active learning stage. In the default configuration, Astronomaly employs the `scikit-learn` implementation of Isolation Forest, although alternative algorithms such as Local Outlier Factor (Breunig et al., 2000) are also available. Objects are assigned anomaly scores and ranked according to their degree of isolation within the feature space.

A key feature of Astronomaly is its personalised active learning approach. Rather than relying solely on anomaly scores, the user assigns relevance scores to a small number of objects using a scale from 0 to 5 based on their scientific interests. These labelled examples are then used to train a RF Regressor (Breiman, 2001), which predicts relevance scores for the remaining unlabeled objects. These predictions are adjusted according to their proximity of previously labelled objects in the feature space. This allows Astronomaly to prioritise objects predicted to be interesting while retaining the ability to explore less well-sampled regions of the feature space (Lochner and Bassett, 2021).

For light-curve datasets, in-built feature extraction is provided by the `feets` (Cabral et al., 2018) package, after which the resulting feature vectors are passed to the anomaly detection and active learning stages. For images and spectra, feature extraction methods such as wavelet decomposition, ellipse-fitting and Fourier based methods are available. The iterative feedback process enables Astronomaly to refine recommendations using only a small number of user-provided labels while maintaining sensitivity to previously unknown classes of anomalies.

### *2.3.* Protege

Protege is an extension of the original Astronomaly framework that replaces the combination of anomaly detection and subsequent active learning with a sequential recommendation approach based directly on user-provided labels (Lochner and Rudnick, 2025). It is implemented within the same Python backend and JavaScript web interface framework in which users can visualise data, inspect objects, and provide labels through an interactive environment. Unlike Astronomaly which begins with an anomaly detection stage and subsequently learns user preferences, Protege does not require an initial anomaly score or ranking. Instead, it operates directly on a set of feature vectors and can be initialised using any user-defined ordering or object selection strategy.

During an interactive session, the user labels candidate objects according to their level of interest (ranging from 0–5, as in Astronomaly). These labels are used to train a Gaussian Process (GP, Rasmussen and Williams 2006) Regressor, which predicts the expected relevance of unlabelled objects and provides an estimate of the uncertainty associated with those predictions. This GP model is then used to select new candidate objects for inspection. In this selection, objects with high predicted relevance are prioritised, and objects with high predictive uncertainty are sampled to improve coverage of the feature space. Following each round of user feedback, the GP model is updated and a new set of candidate objects is proposed.

[1] The user may also choose not to label the presented object.

[2] `https://coniferest.snad.space/`

[3] A comparison of Active Anomaly Discovery and PineForest is present in an upcoming PineForest paper (Ishida et al., in preparation) therefore we have deemed it beyond the scope of this work.

## 3. Light-Curve Analysis using ZTF Data

We first evaluate the three A-AD frameworks using the time-domain light curve data as a relatively lower dimensional test case. This section describes the construction of the dataset, the extraction and selection of the features used to represent it, and the experiments performed to compare the behaviour of the algorithms.

### 3.1. ZTF Light-Curve Dataset

The light curve dataset used in this work (D1 hereafter) consists of r-band light curves in field 327 from the Zwicky Transient Facility (ZTF, Bellm et al. 2019) database through the IRSA system[4] which returns the latest version (`DR20` when this work was started) of the light curves available. ZTF conducts wide-field scans of the optical transient sky ($g$,$r$ and $i$ bands) to detect, characterise and measure parameters of objects as diverse as supernovae to more commonplace objects such as asteroids passing by the earth using a custom-built mosaic camera mounted on the Samuel Oschin Telescope at Palomar Observatory and has been operational since 2018. The initial dataset contained 826, 252 objects. Once the light curves were obtained, we performed some cuts to better define the dataset. We removed duplicate observations and objects with fewer than 30 observations. A further quality cut was made by retaining only those objects with $catflags = 0$ which is a ZTF measure corresponding to epochs with no detected data or image quality issues[5]. After all the cuts were made, there were 430, 431 remaining objects. This is the dataset used for the light curve experiments in this work.

### 3.2. Feature Extraction

Feature extraction is an important step in many machine-learning tasks, reducing the complexity and dimensionality of the input data and thereby making subsequent analysis less computationally intensive. The choice of representation is particularly important, as the extracted features must retain the information required to distinguish between different classes of objects. Approaches commonly used for astronomical data include conventional dimensionality-reduction techniques such as Principal Component Analysis (PCA; Pearson 1901), model-based methods in which descriptive parameters are derived from fitted functions, and more recent deep-learning approaches that utilise latent representations learned by neural networks.

A commonly used feature extraction approach for light curves is to obtain descriptive features that embody the shape and variability of the light curve itself which can be used to characterise the object/object class to which it belongs. The feature extraction package used in this work for D1 is `light-curve` (Malanchev et al. 2021; Lavrukhina and Malanchev 2023)[6], time series package developed by the SNAD team[7] as part of pipeline development. It is described in more detail in the following section.

### 3.3. light-curve

`light-curve` is a feature extraction package for irregular time-series data that was developed as part of SNAD pipeline developments as a mixed Rust/Python package, that provides a Python interface for Rust-written feature extractors, as well as additional pure Python implementations of feature extractors which are used for cross-testing and fast probing of some experimental features. Here, we utilise high performance Rust implementations of feature extractors through the provided Python interface which extends the experimental scope.

For single light curves, it takes as input, time series data such as magnitude and time units and returns features that trace magnitude and summary statistics for periodic signals which describe different characteristics of the shape of the light curve. In this work, we use magnitude, magnitude error and MJD (Modified JUlian Date). For multiple light curves, there is a Rust multithreading option (`.many()`); for D1, feature extraction for the full dataset of 430,431 light curves took 3.9 minutes. There is also support for feature extraction from multi-band light curves. Among implemented features, there are several based on parametric fit functions (Bazin fit (Bazin et al., 2009), Rainbow (Russeil et al., 2024) & Villar (Villar et al., 2019)). Moreover, there are two meta-features such as the Lomb-Scargle periodogram (Lomb 1976; Scargle 1982) and time series binning, which can take other feature extractors as input and apply them to pre-processed time series.

### 3.4. Feature Selection

The initial feature set that we have extracted using `light-curve` is recreated following Pruzhinskaya et al. (2023) and contains 42 features to which PCA was applied to reduce the dimensionality of the feature representations and choose the most useful features for A-AD.

PCA represents a dataset containing multiple variables in terms of a smaller number of orthogonal linear combinations of the original variables, known as Principal Components (PCs). These components describe directions of decreasing variance in the data, from *PC1* to *PCn*. Although dimensionality reduction generally entails some loss of information, PCA provides a means of retaining the dominant variance while reducing the number of dimensions.

Prior to PCA, the initial feature sets were standardised using the `StandardScaler` function from `scikit-learn` (Pedregosa et al., 2011), which centres each feature to zero mean and scales it to unit variance. This prevents features with larger numerical ranges from disproportionately influencing the principal components. For D1, the number of components was adjusted to retain 85% of the variance present in the feature representation by fitting PCA with 11 PCs. For each retained principal component, the original feature with the largest absolute loading was selected for subsequent analysis. The choice of the variance threshold was guided by preliminary experiments. For D1, the experiments showed that 85% of the variance provided a compact representation and preserved discriminative information.

[4] `https://irsa.ipac.caltech.edu/`

[5] See the ZTF Science Data System (ZSDS) Explanatory Supplement, Section 10.3 (https://caltech.edu).

[6] `https://pypi.org/project/light-curve/`

[7] `https://snad.space/`

| Chosen light curve parameters for A-AD |
|---|
| chi-squared |
| kurtosis |
| linear_fit_slope |
| linear_trend_noise |
| magnitude_percentage_ratio_40_5 |
| period_1 |
| period_2 |
| period_s_to_n_1 |
| periodogram_amplitude |
| periodogram_beyond_2_std |
| periodogram_cusum |

Table 1: Table showing the `light-curve` features that were chosen for A-AD after performing PCA. The feature definitions are given in Section 3.4.1.

### *3.4.1. Features*

The 11 light curve features selected as a result of PCA are shown in Table 1. Several of these features have the usual meaning as they pertain to any curve. For instance, *chi-squared* is the reduced $\chi^2$ of a linear fit to the light curve, *kurtosis* the kurtosis (tailedness) of the magnitude distribution of the light curve, *linear_fit_slope* the slope of a linear fit to the light curve and *linear_trend_noise* its standard deviation. *magnitude_percentage_ratio_40_5* calculates the ratio of the $40^{th}$ and $5^{th}$ inter-percentile ranges for magnitude values. The rest of the features relate to the periodogram fit (Lomb-Scargle) to the light curve data. *period_1* is the period of a spectral curve with 2 peaks, *period_2* that of one fitted with 3 peaks & *period_s_to_n_1* the signal-to-noise ratio of a periodogram spectral fit with one peak. *periodogram_amplitude* is the half-amplitude of the periodogram, *periodogram_beyond_2_std* calculates the fraction of values in the periodogram above or below two standard deviations from the weighted average and *periodogram_cusum* the range of the cumulative sums of the periodogram values.

### *3.5.* A-AD *Implementation*

PineForest was implemented using the coniferest package with 256 active trees (n_trees=256) which form the forest used to calculate the anomaly ranking, and 768 spare trees available for replacement during the tree-filtering process (n_spare_trees=768), using a random seed of 42 to ensure reproducibility. The numbers of active and spare trees follow the configuration provided in the `coniferest` documentation; preliminary tests with moderately different values produced no substantial change in the results. User interaction was handled through the viewer_decision_callback, which opened each selected candidate in `SNAD Viewer` (Malanchev et al., 2023) for visual inspection and label assignment. Each A-AD was terminated after 100 user decisions using the TerminateAfter() callback, providing a common labelling budget for comparison between the three frameworks.

In the Astronomaly implementation, D1 was loaded using the LightCurveDataset class, with time (MJD), magnitude, magnitude error, and object identifier (ZTF OID) columns specified. The precomputed feature vectors were scaled using FeatureScaler before anomaly detection. As explained in Section 2.2, the initial anomaly ranking was obtained using Isolation Forest. To ensure a fair comparison with PineForest, the number of trees in the Isolation Forest was set to 1024, matching the total number of active and spare trees used in the PineForest configuration, and the objects were arranged in the decreasing order of anomaly scores.

A-AD was performed using NeighbourScore, with one object labelled per iteration and the model retrained after each label. The resulting scores were converted to the 0–5 Astronomaly score range using ScoreConverter and sorted in decreasing order of anomaly score again. The RF Regressor used during active learning was also configured with 1024 trees. For visualisation only, the scaled light-curve feature space was projected using t-SNE with a perplexity of 100 on the JavaScript interface. Candidate objects were visually inspected through this interface, assigned relevance scores, and the A-AD session was continued for 100 user-feedback iterations.

For the Protege implementation, we used the same input data and feature representations as in the corresponding Astronomaly runs. Although Protege does not require an initial anomaly detection ranking, we initialised the first candidate ordering using Isolation Forest for consistency with the Astronomaly runs and to avoid an arbitrary ordering based on object identifiers. The Isolation Forest was run with 1024 trees, providing the same initial ranking as for Astronomaly. Following the user labels, the scores were converted to the 0–5 Astronomaly score range using ScoreConverter. These scores were used only to define the initial ordering before GP-based A-AD was performed using the Protege Gaussian Process module with ei_tradeoff=3, which controls the balance between predicted relevance and uncertainty during candidate selection. As for Astronomaly, the light-curve feature space was projected using t-SNE with a perplexity of 100. As with the other methods, one object was labelled per iteration and each session consisted of 100 user-feedback iterations.

### *3.6. Experiments*

The three algorithms that we are comparing here differ substantially in their underlying methodologies, learning strategies and implementations. Therefore a single evaluation metric is insufficient to effectively characterise their respective strengths and limitations. Since expert inspection is the limiting resource in A-AD, performance is typically evaluated under a finite labelling budget rather than allowing the active-learning process to continue indefinitely. Previous astronomical applications have similarly adopted fixed inspection budgets, including evaluation over the first 100 user-labelled objects with Astronomaly (Lochner and Bassett 2021; Etsebeth et al. 2024) and a budget of 50 iterations per field with PineForest (Sreejith et al., 2026).

Table 2: Summary of the comparative experiments performed on the light-curve (D1) and image (D2) datasets, showing the A-AD frameworks, prior information, and purpose of each experiment.

| Experiment | Dataset | Methods | Prior information | Purpose |
|---|---|---|---|---|
| Baseline Comparison | D1 | PineForest, Astronomaly, Protege | None | Baseline behaviour |
| Cross-Method Prior Transfer | D1 | PineForest, Astronomaly, Protege | Method-generated labels | Effect of prior transfer |
| Targeted-Prior Experiment | D1 | PineForest, Astronomaly, Protege | 15 RR Lyrae | Effect of targeted prior |
| Baseline Comparison | D2 | PineForest, Astronomaly, Protege | None | Baseline behaviour |
| Cross-Method Prior Transfer | D2 | PineForest, Astronomaly, Protege | Method-generated labels | Effect of prior transfer |

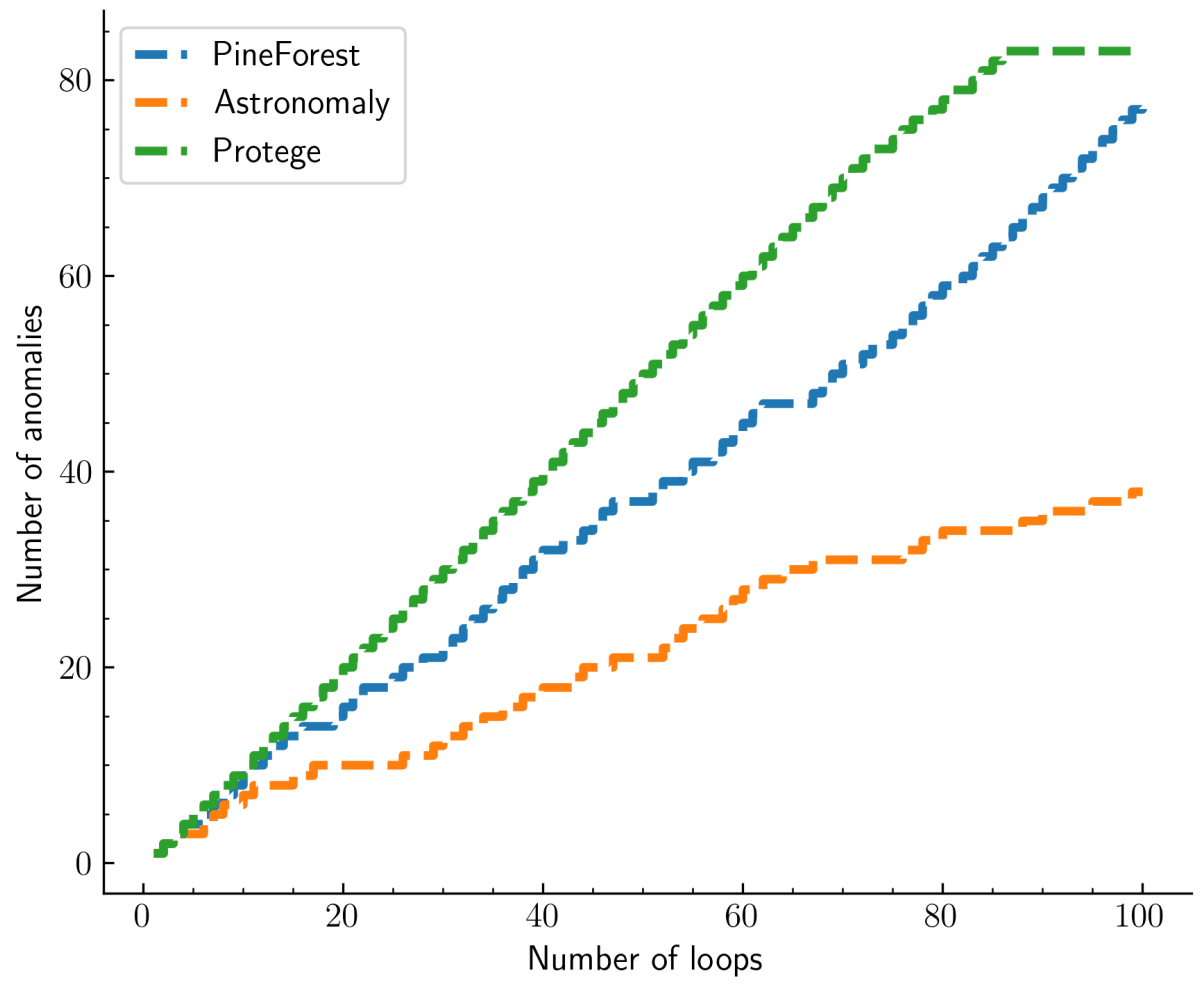

Figure 1: Cumulative number of anomalies identified during the Baseline Comparison (Section 3.6.1) on the light-curve dataset (D1). PineForest, Astronomaly and Protege were all initialised without prior labels, and retrained in each loop after providing user feedback. Protege recovers anomalies at the highest rate throughout most of the experiment, followed by PineForest, while Astronomaly identifies substantially fewer anomalies under the same experimental conditions.

We therefore adopt a common budget of 100 iterations for each method, allowing us to compare how rapidly the three frameworks adapt to the user's definition of an interesting anomaly. We designed and performed three complementary independent experiments, which we describe in the following sections (also described in Table 2).

### 3.6.1. *Baseline comparison*

For the Baseline Comparison between the three A-AD algorithms, "variable stars" were adopted as the anomaly definition. Known variable-star classifications were obtained from the International Variable Star Index (VSX)[8], while objects identified as variable through visual inspection but without a corresponding VSX classification were labelled as VS:. QSO/AGN classifications were obtained from Gaia (Gaia Collaboration et al. 2023; Manteiga et al. 2021). Each algorithm was run without prior labels for the common budget of 100 user-labelled objects. Since PineForest updates after every user decision, Astronomaly and Protege were also retrained after each label in this experiment, ensuring that all three methods received the same amount and frequency of user feedback.

Figure 1 shows the cumulative recovery of variable stars over the 100 iterations, with the slope of each curve representing the rate of anomaly recovery. Steeper sections indicate that a larger fraction of the objects being recommended are classified as anomalies. These curves are not expected to converge within the fixed 100 iterations budget, rather, their evolution shows whether each method continues to prioritise anomalies as user feedback accumulates. Protege recovered the largest number (83), followed by PineForest (78) and Astronomaly (38). The three methods perform similarly during the first ~10–15 iterations, after which Protege and PineForest recover variable stars at a substantially higher rate than Astronomaly.

The composition of the inspected samples also differed substantially (Figure 2). In addition to 78 variable stars, PineForest returned 20 artefacts, one QSO/AGN and one normal star; Astronomaly returned 38 variable stars, 54 artefacts, six QSO/AGN and two stars; and Protege returned 83 variable stars, 14 artefacts, two QSO/AGN and one star. Of the recovered variable stars, 16, 9 and 3 were previously uncatalogued variable-star candidates for PineForest, Astronomaly and Protege, respectively.

To assess the learned rankings after the active-learning phase, we additionally classified the next 100 objects recommended by each algorithm without using these labels for further retraining (Figure 3). Protege remained concentrated on variable stars, returning 74 known variables and 12 uncatalogued variable-star candidates, compared with nine artefacts, two QSO/AGN and three non-variable stars. PineForest returned 33 known variables and 23 uncatalogued candidates, together with 20 artefacts, 14 QSO/AGN and 10 stars. In contrast, 97 of the 100 objects recommended by Astronomaly were artefacts, with the remaining three being uncatalogued variable-star candidates.

The marked differences in these post-training recommendations indicate that, despite receiving the same amount of user feedback, the three A-AD strategies converge towards substantially different regions of the feature space. The fixed 100-label budget does not imply that any of the methods has converged, however, and additional supervision may further modify the learned rankings. We investigate this in the following experiment by supplying labelled examples as prior information.

### 3.6.2. *Cross Method Prior Transfer*

Semenikhin et al. (2026) showed that PineForest responds positively to prior knowledge by pruning trees that are inconsistent with the supplied anomaly definition. Here, we investigate

[8] https://vsx.aavso.org/

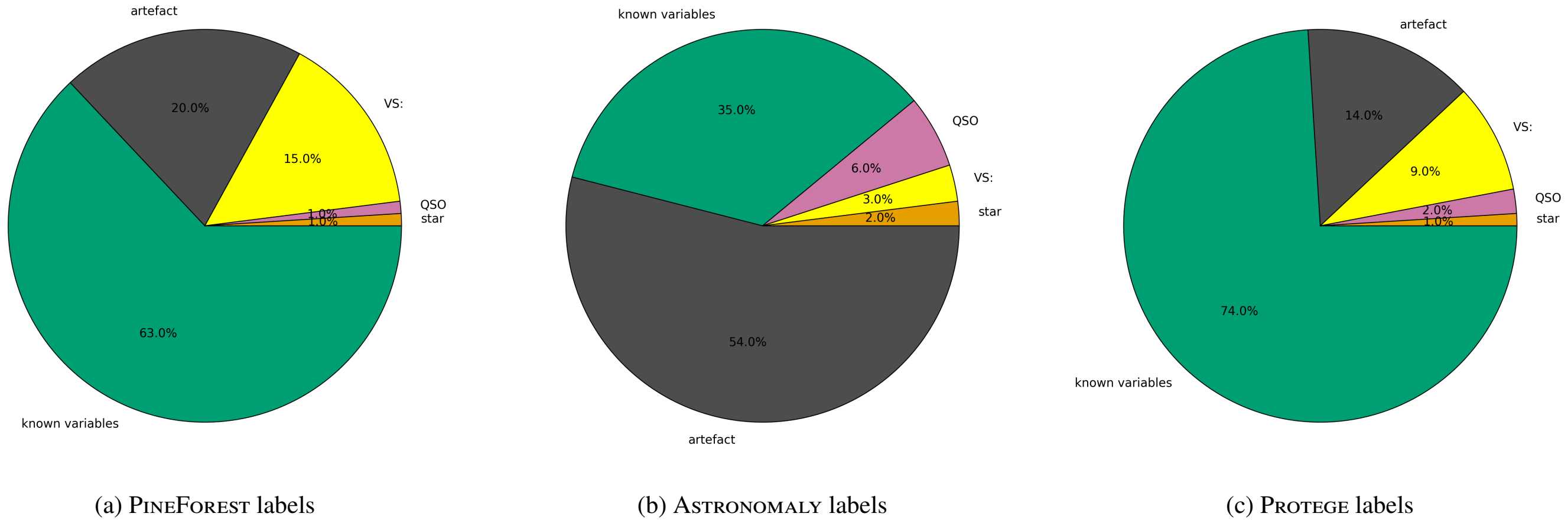

(a) PineForest labels (b) Astronomaly labels (c) Protege labels

Figure 2: Distribution of object labels in the initial run for all algorithms. The pie charts show the composition of the first 100 objects identified by PineForest, Astronomaly, and Protege during the initial active learning experiment on D1.

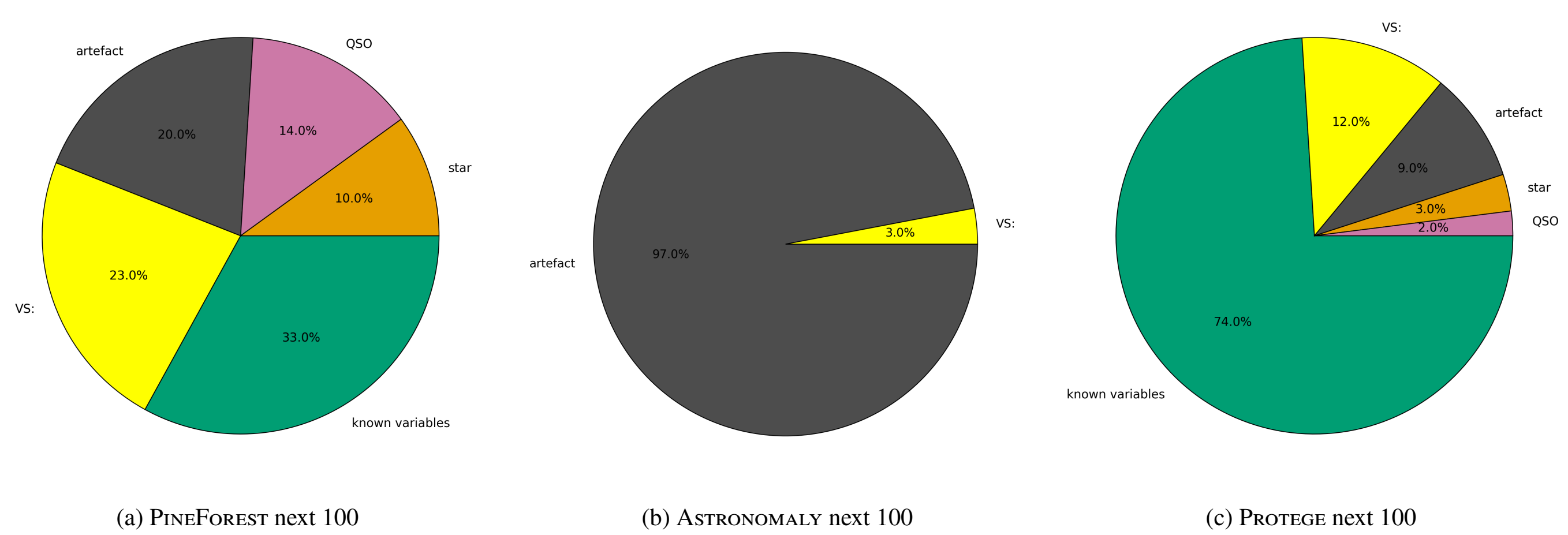

(a) PineForest next 100 (b) Astronomaly next 100 (c) Protege next 100

Figure 3: Distribution of object labels in the subsequent 100 objects for all algorithms. The pie charts show the composition of objects identified by PineForest, Astronomaly and Protege for the 100 objects after A-AD is completed on D1.

the effect of prior information across all three A-AD frameworks.

Instead of constructing a separate labelled dataset, we used the first 100 labelled objects from the Baseline Comparison (Section 3.6.1) as three independent prior datasets, corresponding to the labels generated by PineForest, Astronomaly and Protege. Each prior set was supplied to all three algorithms before a new A-AD session, resulting in nine combinations of algorithm and prior dataset. This allows us to distinguish the effect of the A-AD strategy from that of the composition of the prior labels. Figure 4 summarises the main results of this experiment.

The response to the addition of prior labels differs between the three algorithms. Protege shows consistently high recovery across all three prior sets, identifying 80, 88 and 88 variable stars when initialised with PineForest, Astronomaly and Protege priors, respectively. This indicates that Protege's GP based recommendation strategy is relatively insensitive to the source of the initial labels.

PineForest is more dependent on the supplied prior dataset. When initialised using its own labels or those of Protege, PineForest recovered 54 and 56 variable stars respectively, while using Astronomaly priors resulted in the recovery of a smaller number of variable stars (50). Although the total variable-star recovery changes only modestly, the composition of its recommendations is more sensitive to the prior set, with Astronomaly-derived priors producing a larger fraction of QSO/AGN.

Astronomaly exhibits the smallest variation, recovering between 14 and 16 variables regardless of label source (PineForest 14, Protege 15, Astronomaly 16) with its recommendations remaining dominated by artefacts regardless of the source of the prior labels.

Taken together, these results indicate that the influence of priors is algorithm dependent, with the choice of A-AD strategy having a larger effect on anomaly recovery than the source of the prior labels for D1. The variable star candidates (VS:) obtained from this experiment and the Baseline Comparison (Section 3.6.1) are set aside to be examined and posted to VSX.

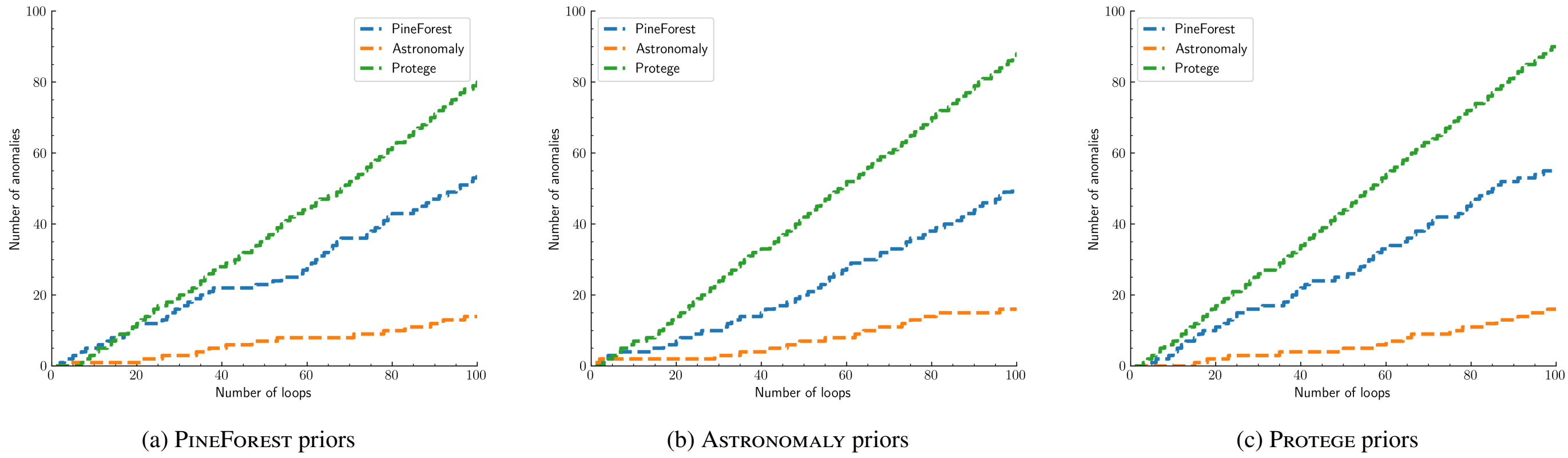

(a) PineForest priors

(b) Astronomaly priors

(c) Protege priors

Figure 4: Cumulative number of anomalies recovered by all three algorithms when the labels from the Baseline Comparison (Section 3.6.1) for D1 are supplied as prior information. As in Section 3.6.1, Protege achieves the highest anomaly recovery, followed by PineForest, while Astronomaly identifies considerably fewer anomalies on D1.

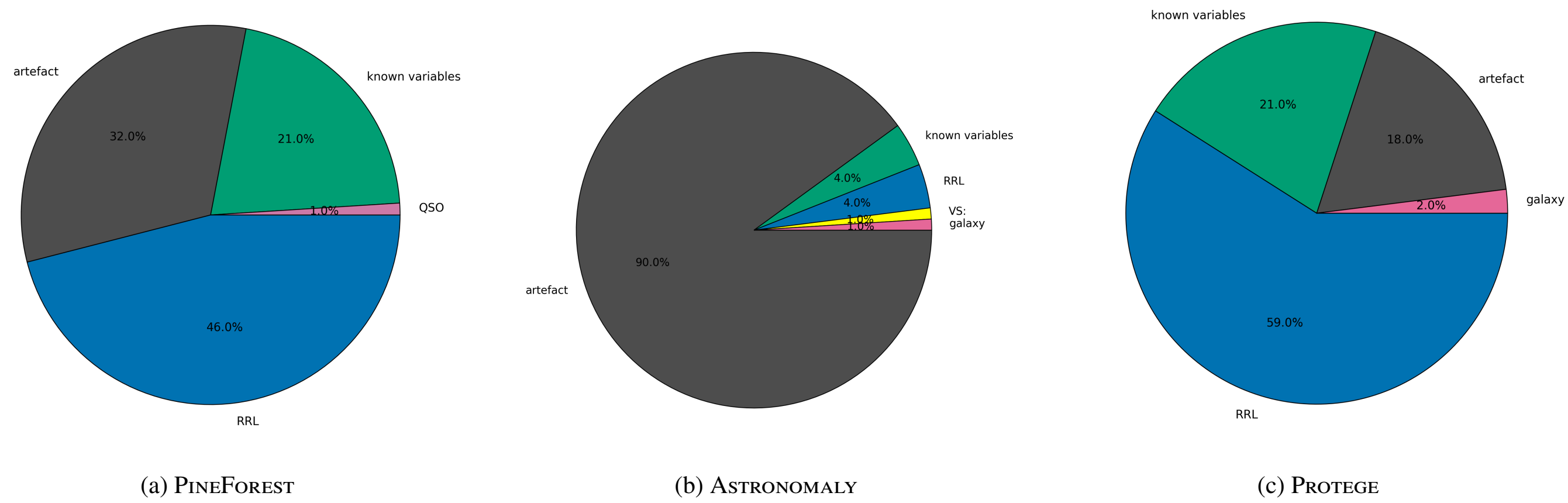

(a) PineForest

(b) Astronomaly

(c) Protege

Figure 5: Distribution of labels in the best case scenario for all algorithms. 15 RR Lyrae stars are given as priors and each algorithm is run with their recommended settings as described in Section 3.6.3.

### *3.6.3. Targeted Prior Experiment*

The previous experiments (Sections 3.6.1-3.6.2) imposed a common iterative protocol to facilitate direct comparison, whereas the three frameworks are normally used with different feedback strategies. We therefore performed a Targeted Prior Experiment in which each method was used according to its intended A-AD workflow. All three were initialised with the same prior set of 15 VSX-confirmed RR Lyrae (RRL) stars. PineForest incorporated these priors and continued to update after each of the 100 user decisions, while for Astronomaly and Protege the priors were loaded, 100 objects were labelled in bulk, and the RF and GP models were then retrained. We therefore compare the first 100 objects encountered during the PineForest A-AD session with the first 100 objects in the post-retraining rankings of Astronomaly and Protege.

The resulting object distributions for these are shown in Figure 5. Protege produced the strongest concentration of the targeted class: 59% of its first 100 post-retraining recommendations were RRL stars, with a further 21% comprising other known variables; the remaining objects were 18% artefacts and 2% galaxies. PineForest recovered 46% RRL stars and a further 21% other variable stars, together with 32% artefacts and 1% QSO/AGN. In contrast, the Astronomaly ranking remained dominated by artefacts (90%), with only 4% RRL stars, 4% other known variables, and 1% each of an uncatalogued variable-star candidate and a galaxy. Since Astronomaly and Protege began from the same Isolation Forest ranking, their different post-retraining compositions demonstrate that their respective A-AD strategies respond differently to the same targeted supervision.

Figure 6 provides a complementary view of how the targeted priors affect the rankings produced by each algorithm. We randomly selected a subset of 50 VSX-confirmed RR Lyrae stars from D1 that were not included among the 15 prior objects and compared their positions before and after the Targeted Prior Experiment. Figure 6 shows the cumulative fraction of these target RRLs recovered as a function of absolute rank, with lower ranks corresponding to objects prioritised earlier for inspection.

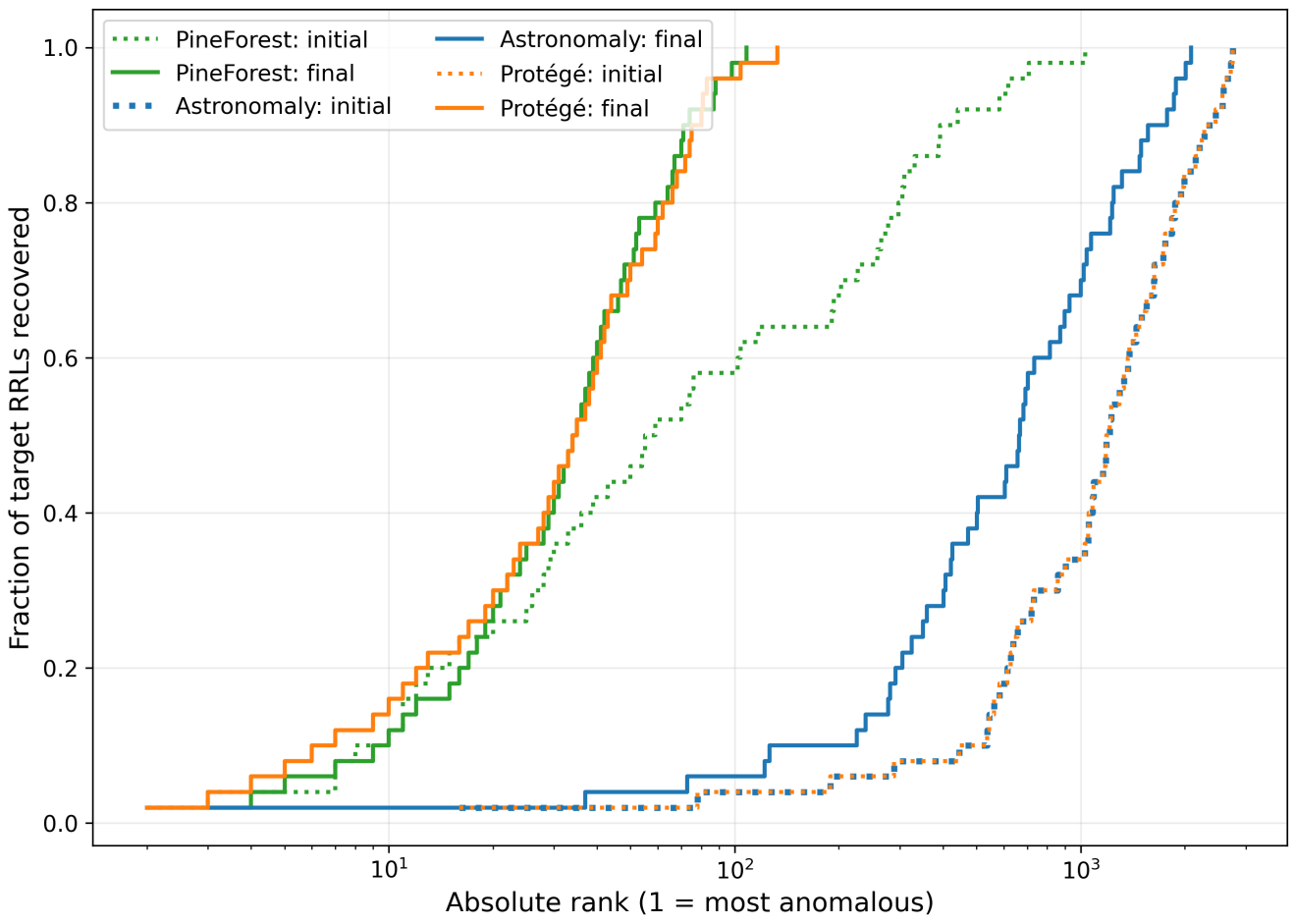

Figure 6: Cumulative absolute-rank distributions of the 50 target RR Lyrae (RRL) objects before and after incorporating the targeted RRL priors for Astronomaly, Protege, and PineForest. The curves show the fraction of target RRLs recovered within a given absolute rank, where lower ranks correspond to objects prioritised earlier for inspection. Dashed and solid lines indicate the initial and final rankings, respectively. The absolute rank is shown on a logarithmic scale.

The initial rankings reveal substantial differences between the methods. The median initial rank of the target RRLs was 57 in the initial `coniferest` Isolation Forest ranking used by PineForest, compared to 1199 for both Astronomaly and Protege (since they both start from the same initial `scikit-learn` Isolation Forest). Following A-AD with the targeted priors, the median ranks were 34.5, 665.5 and 34.5 for PineForest, Astronomaly and Protege, respectively. This difference is also evident when considering the fraction of target RRLs ranked within the top 100. In the initial rankings, 58% of the target RRLs were within the top 100 for PineForest, compared with only 4% for both Astronomaly and Protege. Following A-AD, these fractions increased to 98%, 6% and 96% for PineForest, Astronomaly and Protege, respectively. The relatively small change for PineForest reflects the stronger starting ranking provided by the `coniferest` Isolation Forest, whereas Protege shows the largest response to the targeted priors and ultimately achieves a concentration of target RRLs within the top 100 comparable to PineForest.

## 4. Image Analysis

In this section we describe the evaluation of the three A-AD frameworks using higher-dimensional image data in experiments similar to Section 3.

### 4.1. Euclid Image Data

The image dataset used in this work (D2 hereafter) is drawn from the Euclid Q1 (Quick Data Release, Euclid Collaboration et al. 2026a) Morphological Catalogue (Walmsley et al., 2025), which provides morphological classifications for approximately 378,000 sources from the Euclid Q1 Data Release. The morphological classifications were obtained using a combination of volunteer annotations from Galaxy Zoo [9] and machine learning predictions based on Zoobot foundation models (Walmsley et al., 2022). Composite 3-colour images (VIS-J-Y) (Euclid Collaboration et al. 2026h; Euclid Collaboration and et al. 2026) for these sources were created using the Euclid Legacy Bulk Cutout Pipeline [10].

To remove contaminants/artefacts and poorly resolved objects, we performed quality cuts on both the catalogue level and after visual examination of the images. We followed the catalogue-level cuts used in the source selection of Xu et al. 2026. We required `VIS_DET=1`, `SEGMENTATION_AREA`$\geq$ 259, `MUMAX_MINUS_MAG`$\geq -2.6$, `MU_MAX`$\geq$ 15.0, and `SPURIOUS_PROB`$<$ 0.05. These criteria preferentially retain well-resolved VIS detections while reducing contamination from stars, saturated sources and spurious detections. In addition, we rejected sources with `point_like_prob>0.2`, with this threshold chosen empirically following visual examination of the images. We also excluded objects with `det_quality_flag` bits 1, 2, 4 or 8 set, as recommended in the Euclid SGS Data Product Description Document[11], and removed objects with NaN values in the selected catalogue features. This resulted in 326,642 objects left in the catalogue. Upon visual examination, we realised that there were still artefacts present in the catalogue in the form of dichroic ghosts, saturated stars, pixel defects, asteroid trails etc. that were not caught in the earlier cuts. In order to remove these as much as possible, we used a three-pronged method combining UMAP (Uniform Manifold Approximation and Projection, McInnes et al. 2018), an initial Isolation Forest and a real-bogus classifier developed for this work. UMAP was used to project the high-dimensional feature space into two dimensions, allowing regions dominated by morphologically similar artefacts to be identified visually. As this process required the extracted image features, the additional artefact removal was performed after feature extraction as described in Section 4.2.3. Throughout this work, sources are identified using their Euclid OBJECT_ID; these are signed integers derived from the source coordinates, such that sources at negative declination have negative identifiers (see the Euclid Q1 MER Final Catalog documentation).[12]

### 4.2. Feature Extraction and Dataset Refinement

#### 4.2.1. VICReg

Image features were extracted using a self-supervised representation learning approach based on Variance-Invariance-Covariance Regularization (VICReg, Bardes et al. 2022). Unlike supervised methods, VICReg does not require labelled training data. Instead, two augmented views of the same image are generated and passed through a shared neural network. The network is then trained to produce similar representations for the two views while simultaneously ensuring that the learned feature representations remain informative and non-degenerate through invariance, variance, and covariance regularisation.

[9] https://www.zooniverse.org/projects/zookeeper/galaxy-zoo
[10] https://github.com/mwalmsley/bulk-euclid-cutouts
[11] https://euclid.esac.esa.int/dr/q1/dpdd/index.html
[12] https://euclid.esac.esa.int/dr/q1/dpdd/merdpd/dpcards/mer_finalcatalog.html

The backbone network consisted of a `ResNet-18` architecture (He et al., 2016) followed by a three-layer projection head with an output dimension of 2048. During training, two independently augmented views of each input image were created using random horizontal and vertical flips, random rotations up to 180°, resizing to $128 \times 128$ pixels and additive Gaussian noise. The network was trained for 200 epochs using the `AdamW` optimiser (Loshchilov and Hutter, 2019) with a learning rate of $3 \times 10^{-4}$ and a weight decay of $10^{-4}$. This network was implemented using the `PyTorch` framework (Paszke et al., 2019).

Although the VICReg objective was trained using features from the fourth residual block (`layer4`) as input to the projection head, feature extraction was performed using the output of the third residual block (`layer3`) of the ResNet backbone. Preliminary experiments showed that embeddings extracted from `layer4` contained a large fraction of zero-valued feature dimensions, resulting in a less informative representation. In contrast, features extracted from `layer3` exhibited greater variability while still capturing the relevant morphological information, and therefore the `layer3` backbone features were adopted for all subsequent analyses. The backbone features were extracted directly rather than from the projection head, producing a 256-dimensional embedding for each image. These embeddings formed the initial image feature set and were subsequently standardised and reduced using PCA to provide a more compact representation for the A-AD algorithms as explained in Section 4.2.2.

#### *4.2.2. Initial Feature Selection*

After feature extraction (Section 4.2.1), and creating the initial feature set, we applied PCA to reduce the 256-dimensional VICReg embeddings to a more compact representation. Similar dimensionality reduction of learned image representations has been adopted for unsupervised analysis of astronomical images (Mohale and Lochner, 2024), including PCA compression of deep galaxy-image embeddings in Euclid (Walmsley et al., 2022). We retained 25 principal components, corresponding to 98% of the variance, as determined from preliminary experiments. These features were subsequently combined with a subset of Euclid catalogue parameters (Euclid Collaboration et al., 2026g) (Table 3), selected to provide additional physically interpretable information beyond that captured by the learned image embedding, resulting in a feature set of 41 features.

#### *4.2.3. Artefact Removal*

We applied an Isolation Forest algorithm (n_estimators = 500) to this combined dataset. We examined and labelled the top 5% of the objects ranked by their Isolation Forest anomaly scores and used this as the training set for a real-bogus classifier using ExtraTreesClassifier from `scikit-learn` (Pedregosa et al., 2011). After visual examination, objects that had a bogus_score$>$ 0.75 were discarded. To this dataset, we applied UMAP in order to examine the projection and confirm that the artefact removal has been effective. In this step we discovered some saturated stars that had escaped the earlier filters, which were also removed. From the 326,642 objects remaining after the catalogue-level cuts, the additional artefact-removal procedure removed 43,288 objects (13.3%), resulting in a final D2 sample of 283,354 objects. This relatively large reduction reflects the removal of image artefacts and contaminants that would otherwise dominate the subsequent anomaly searches.

#### *4.2.4. Final Feature Set*

The final image feature set used for A-AD consisted of two complementary components: features learned directly from the images using VICReg and physically motivated photometric and morphological parameters obtained from the Euclid catalogue.

The additional Euclid parameters comprised measures of source compactness (*point_like_prob*), total and PSF-fitted fluxes (*flux_detection_total*, *flux_vis_psf*), object size (*kron_radius*, *segmentation_area*, *equiv_radius_from_area*), central light concentration (*mumax_minus_mag*, *concentration*), projected shape (*ellipticity*), and non-parametric morphological descriptors (*asymmetry*, *smoothness*, *gini*, and *moment_20*), which characterise the symmetry, clumpiness, and spatial distribution of the light. Near-infrared colour indices (*y_minus_j*, *j_minus_h*, and *y_minus_h*) derived from the Euclid NISP photometry were also included to characterise the spectral energy distribution of each source.

Together, these catalogue-derived parameters complement the learned image embeddings by incorporating physically interpretable photometric, structural, and colour information into the anomaly detection process. The inclusion of the NISP colours means that the resulting feature representation is not exclusively morphological, but also contains information about the spectral properties of the sources. This allows A-AD to respond to unusual combinations of morphology and colour, rather than identifying anomalies solely on the basis of their visual structure. The NISP colours may therefore provide additional discriminatory information for objects with similar optical morphologies but different stellar populations, dust properties, or redshifts, although the relative contribution of the photometric and morphological features is not separately quantified in this work.

### *4.3.* A-AD *Implementation*

The algorithm configurations described for D1 were retained for the image experiments on D2, including the number of trees used by PineForest and Astronomaly, the initial ranking procedure for Astronomaly and Protege, and the A-AD configurations.

For Astronomaly, images were loaded using ImageThumbnailsDataset, with sigma clipping and scaling applied to normalise the image intensities for visualisation. Only sources with corresponding feature vectors were retained, and the precomputed feature vectors were scaled using FeatureScaler. Unlike the t-SNE projection used for D1, the image feature space was represented in the JavaScript interface using UMAP. Candidate objects were inspected and assigned relevance scores through the same interface.

| Chosen image parameters for A-AD |
|---|
| point_like_prob |
| flux_detection_total ($\mu$Jy) |
| flux_vis_psf ($\mu$Jy) |
| kron_radius (px) |
| segmentation_area (px$^2$) |
| equiv_radius_from_area (px) |
| mumax_minus_mag (mag arcsec$^{-2}$) |
| concentration |
| ellipticity |
| asymmetry |
| smoothness |
| gini |
| moment_20 |
| y_minus_j (mag) |
| j_minus_h (mag) |
| y_minus_h (mag) |

Table 3: Table showing the image features that were chosen for A-AD from the Euclid MER catalogue. The feature definitions are given in Section 4.2.4.

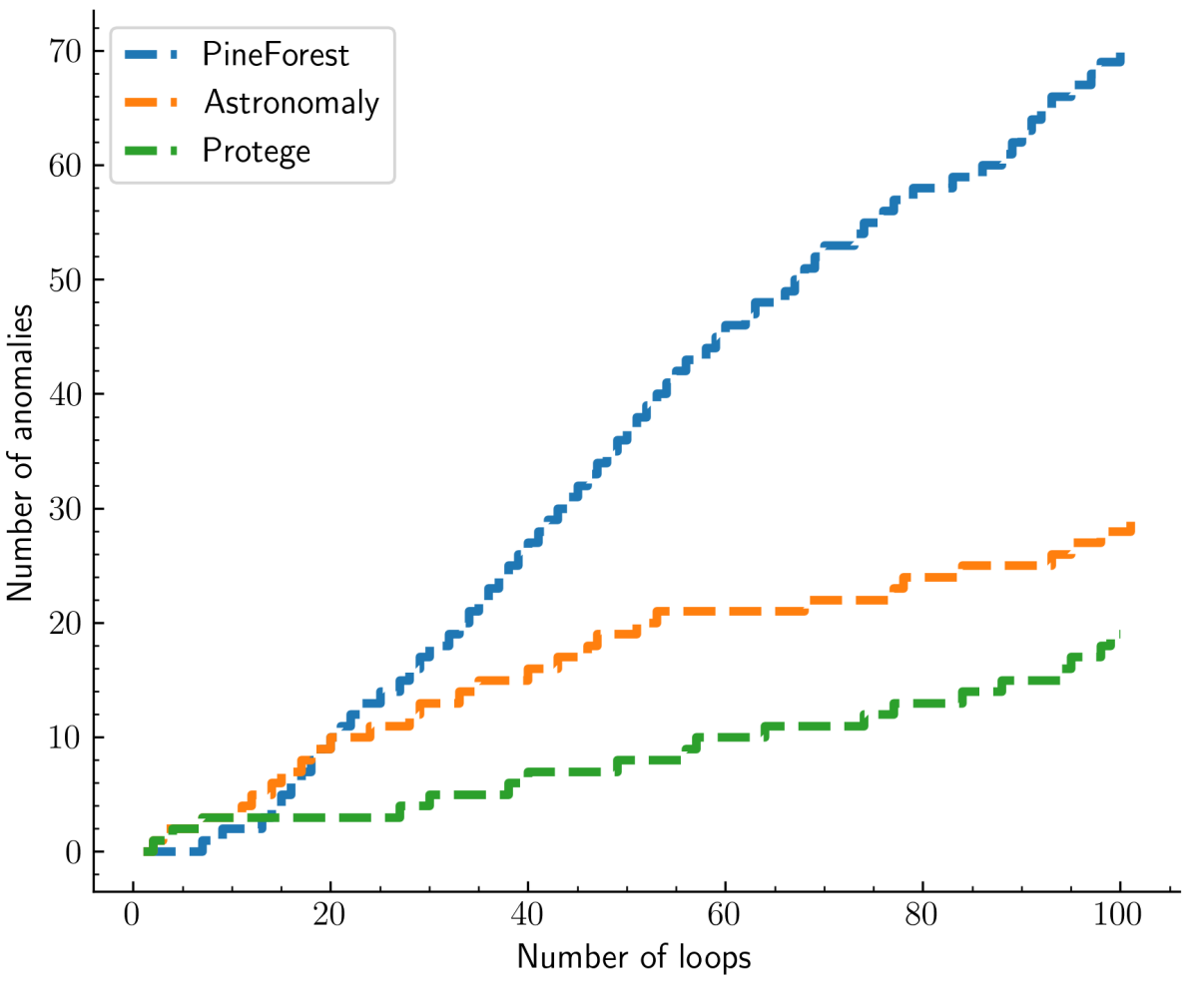

Figure 7: Cumulative number of merger/disturbed galaxies recovered during the first 100 active learning iterations on the image dataset (D2). All three algorithms were initialised without prior labels. PineForest increasingly prioritises the target morphology, substantially outperforming Astronomaly and Protege over the first 100 A-AD iterations.

For Protege, the same D2 input data and feature representations were used as in the corresponding Astronomaly runs. Images were loaded and preprocessed in the same way, and similar to Astronomaly, only image sources with matching feature vectors were retained. The initial candidate ordering and subsequent GP-based A-AD followed the same configuration described for D1.

### *4.4. Image experiments*

#### *4.4.1. Image Baseline Comparison*

The Image Baseline Comparison experiment follows the same protocol as the one described in Section 3.6 for D1 data. Each algorithm was run for 100 iterations without any priors, and each presented object was labelled. For D2, the anomaly definition was 'merging/morphologically disturbed galaxies'. This provides a sufficiently abundant but morphologically specific target class, allowing the A-AD strategies to be compared without the experiment being dominated by a scarcity of positive examples.

Figure 7 shows the results of this experiment for the 100 iterations. PineForest recovered 70 merger/disturbed galaxies during the 100 iterations, compared with 29 for Astronomaly and 19 for Protege. After a slow initial phase, the PineForest curve rises faster, indicating increasing enrichment of the target class as user feedback accumulates, whereas Astronomaly and Protege show lower recovery rates. Thus, while Protege achieved the highest anomaly recovery on D1, the D2 Baseline Comparison favours the tree-filtering A-AD strategy of PineForest. Despite the differences in overall recovery, 10 objects were common to the samples returned by all algorithms, nine of which were visually classified as merger/morphologically disturbed galaxies (Figure 8). These systems show prominent interaction signatures, including tidal features, asymmetric morphologies, overlapping components and multiple nuclei, and represent relatively unambiguous examples of the adopted anomaly class.

As for D1 (Section 3.6.1), we then examined the next 100 objects recommended by each algorithm after the initial A-AD stage without further retraining. PineForest continued to recover the highest fraction of the target class, with 43% of the subsequent 100 objects classified as merger/disturbed galaxies, compared with 23% for Astronomaly and 22% for Protege. The higher target fraction for PineForest therefore persists in its post training ranking.

#### *4.4.2. Image Cross-Method Prior Transfer*

We replicated the Cross-Method Prior Transfer experiment (Section 3.6.2) on D2 using the first 100 labels from the Image Baseline Comparison (Section 4.4.1) as priors for all three methods. Figure 9 shows the cumulative number of merging and morphologically disturbed galaxies identified during the subsequent 100 iterations.

PineForest consistently achieved the highest recovery across the three prior sets, identifying 62, 60 and 54 target galaxies when initialised with PineForest, Astronomaly and Protege priors respectively. The corresponding recoveries for Astronomaly were 17, 11 and 23 while Protege recovered 30, 15 and 22.

Prior information did not universally improve recovery relative to the no prior baseline. PineForest recovered fewer target galaxies than its baseline value of 70 with all three prior sets, although it remained the best performing method. Astronomaly remained below its baseline recovery of 29, while Protege exceeded its baseline value of 19 when initialised with PineForest priors (30) or its own priors (22) but recovered only 15 targets with Astronomaly priors. This may suggest that the GP based recommendation strategy benefits from a more anomaly rich initial labelled set when modelling the target class in the higher-dimensional D2 feature space.

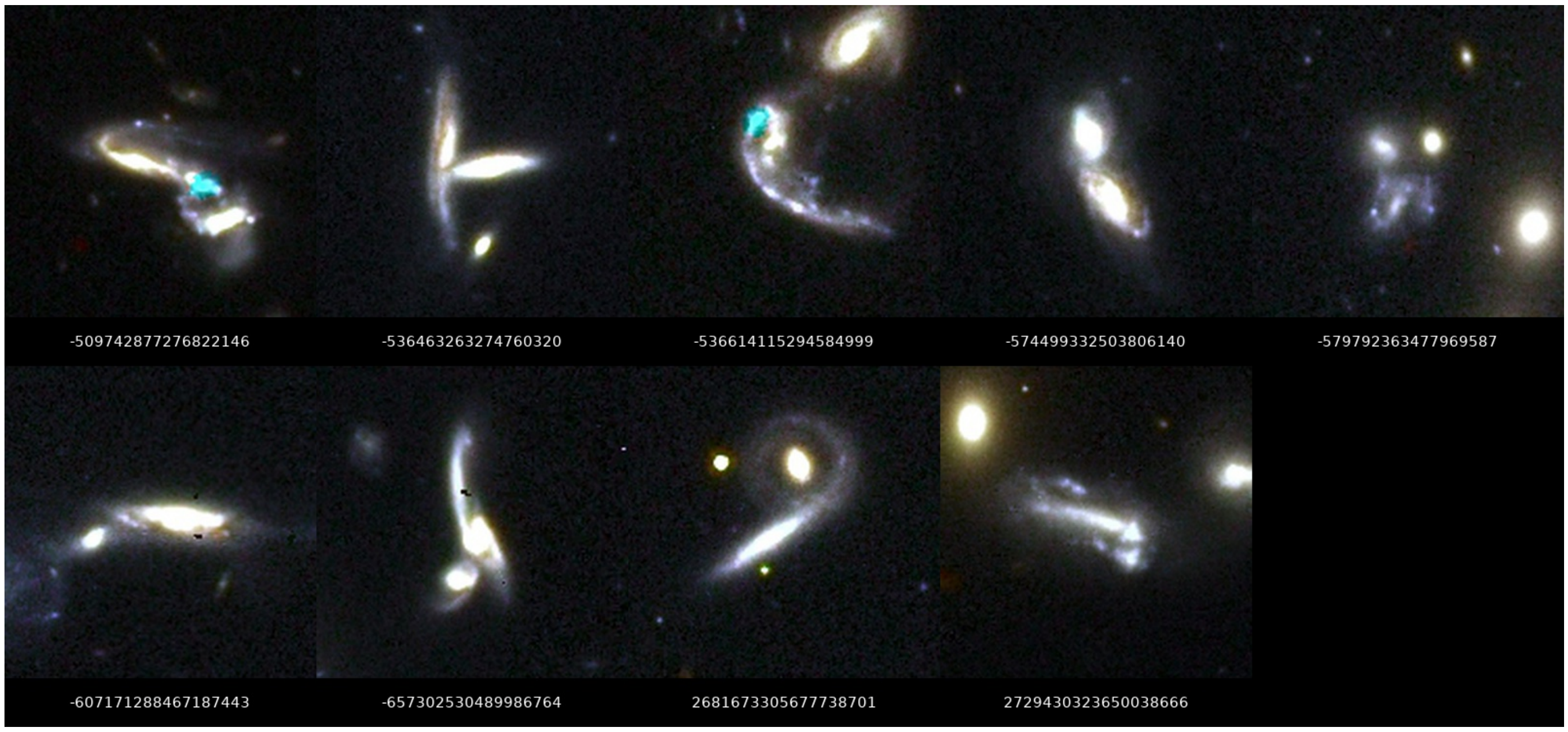

Figure 8: The nine merger/disturbed galaxies recovered by all three algorithms during the Baseline Comparison (Section 4.4.1) on D2. Each object was independently identified by PineForest, Astronomaly and Protege within the first 100 A-AD iterations. The numbers below each image indicate the Euclid object IDs.

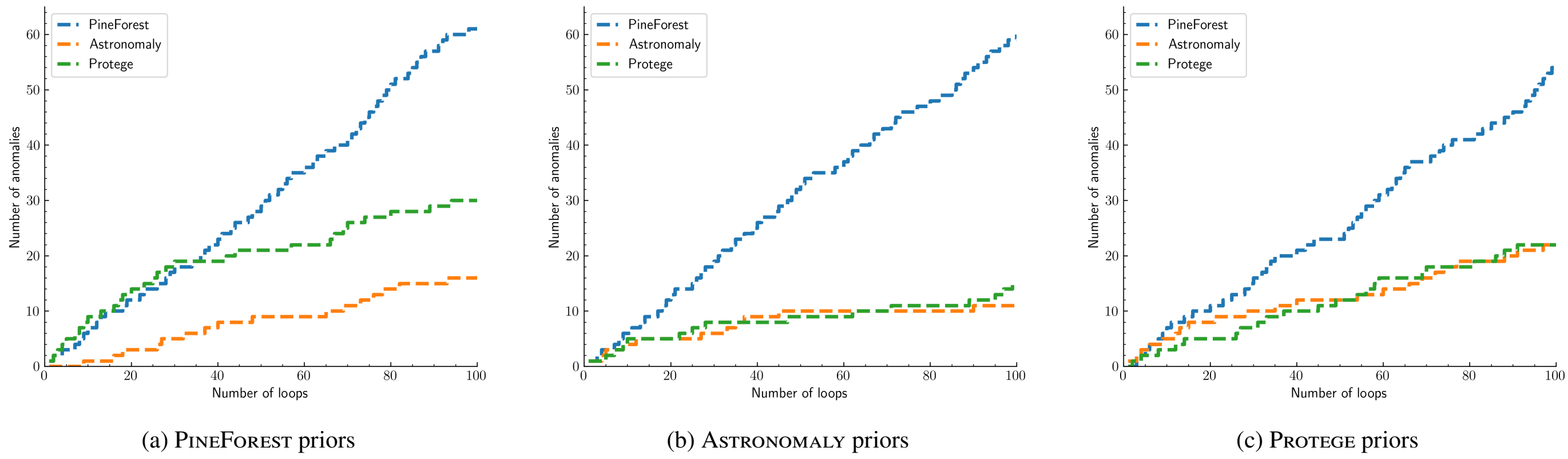

(a) PineForest priors (b) Astronomaly priors (c) Protege priors

Figure 9: Cumulative number of merger/disturbed galaxies recovered during the first 100 A-AD iterations on D2 using prior labels from the Baseline Comparison (Section 4.4.2). Panels show the performance of PineForest, Astronomaly, and Protege when initialised with prior labels obtained from (a) PineForest, (b) Astronomaly, and (c) Protege, respectively.

## 5. Large-scale application to Euclid imaging

### *5.1. Hybrid Workflow*

While the targeted-prior experiment performed on D1 using RR-Lyrae stars as prior information provided a controlled evaluation of A-AD, an equivalent experiment is less meaningful for D2. Unlike variable stars, where a well defined RR-Lyrae subclass can be used to create targeted priors, the Euclid morphological catalogue does not contain a sufficiently large, uniformly labelled sample of specific disturbed galaxy subclasses that can be directly adopted. Even though several Euclid studies have published samples of mergers and other morphological types, these are generally produced using specific selection criteria, visual classifications, or science-specific definitions. This makes them unsuitable as an objective benchmark for A-AD. Therefore, instead of creating priors in that manner, we demonstrate how the complementary strengths of Protege and PineForest can be combined for large scale anomaly discovery.

In order to achieve this, we first performed an extensive A-AD campaign in which SS manually inspected and labelled a total of 10,000 images. The first 1800 were inspected using Protege, a number set by the practical computational limit of Protege on the hardware[13] used in this study. Beyond approximately 1800 labelled objects, GP retraining became unstable and frequently failed due to memory limitations. The resulting labelled cata-logue was then used as prior information for PineForest, and a further 8200 previously unseen images were inspected within one run with that budget.

[13]The analysis was performed on a laptop equipped with a 13th Gen Intel Core i9-13900H processor, 32 GB of system memory, and an NVIDIA RTX 2000 Ada Generation Laptop GPU with 8 GB of GPU memory.

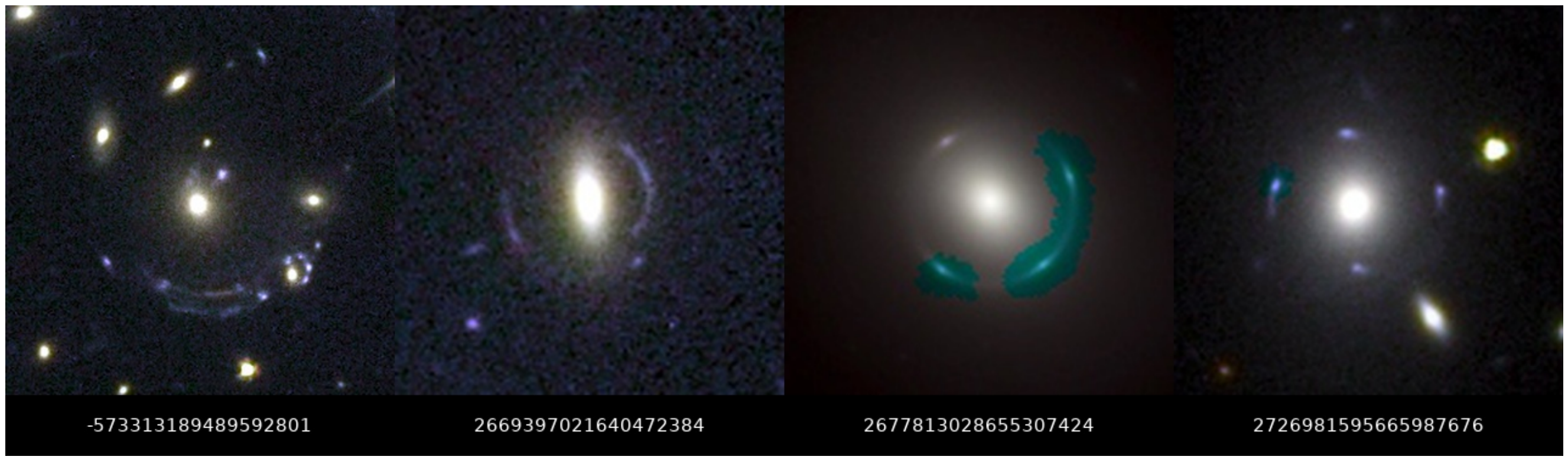

(a) convincing lenses

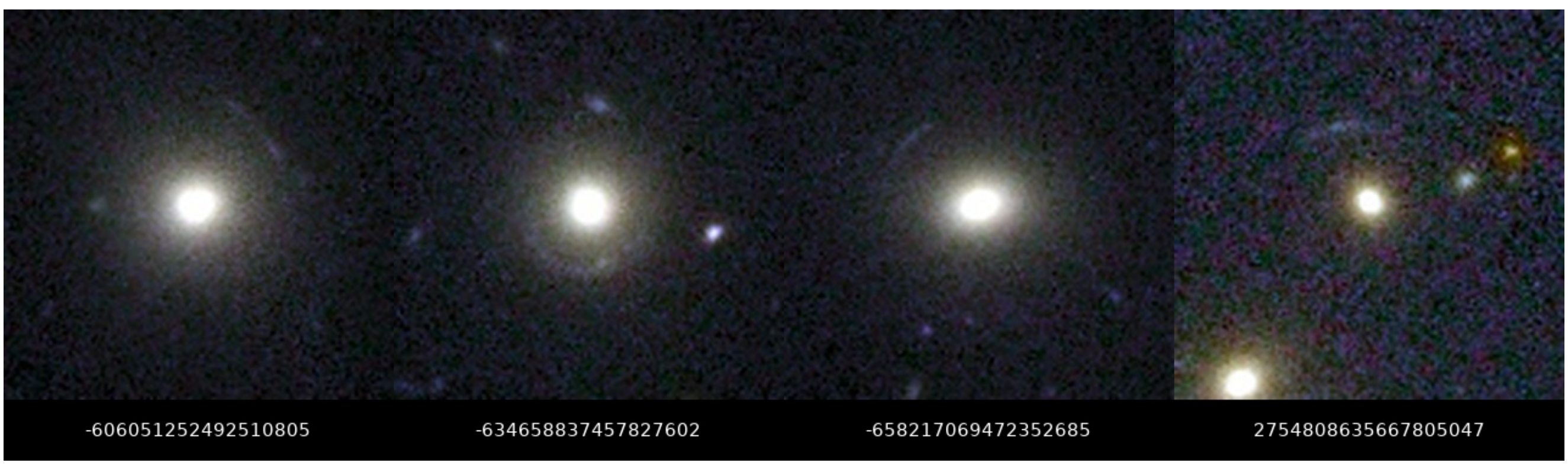

(b) candidate lenses

Figure 10: Representative examples of the strong lens systems recovered by the hybrid active-learning search. (a) Four visually convincing strong lenses, including systems exhibiting complete Einstein rings, partial arcs, and multiple-lens configurations. (b) Four representative candidate strong lenses. Following detailed visual inspection and reclassification, the final sample comprises 12 convincing strong lenses and 17 candidates.

This combined Protege – PineForest hybrid search resulted in 799 candidate anomalies (morphologically interesting objects), which were organised into eight categories based on visual inspection. These candidates were subsequently subjected to a second, more detailed visual inspection by SS and RN. Objects were retained, reassigned to a more appropriate category, or rejected where the apparent anomaly could be attributed to projection or superposition effects, neighbouring sources, or otherwise normal morphological variation. The eight morphological categories were not defined in advance of the A-AD search, but were constructed by SS during the initial visual inspection to group the principal types of unusual morphology represented among the recovered objects. These categories are not always mutually exclusive, because some objects exhibit multiple unusual features, but they provide a convenient framework for summarising the diversity of recovered systems. In the following subsection (Section 5.2), we briefly describe each category.

### *5.2. Morphological categories*

**Strong lenses:** Of the 799 morphologically unusual objects recovered by the hybrid search, 35 were initially classified as potential strong lenses during the first round of visual inspection. Following the detailed visual inspection, they were categorised as 12 convincing, 13 candidates and 10 rejected, based on appearance. Convincing systems showed clear lens signatures, whereas candidate systems exhibit plausible albeit less convincing or low surface brightness features. In addition, 4 objects initially classified as ring-galaxy candidates were reclassified as strong lens candidates following detailed visual inspection (Figure 12). Including these systems, the final strong lens sample therefore comprises 29 objects: 12 visually convincing systems and 17 candidates. We cross-matched the 29 retained strong lens systems and candidates against the published Euclid Q1 strong lens catalogues, including the primary Strong Lensing Discovery Engine catalogue (Euclid Collaboration et al., 2026j), the high-velocity-dispersion search of Euclid Collaboration et al. 2025, the additional bright and low redshift systems identified by Euclid Collaboration et al. 2026d, the AstroVink search (Euclid Collaboration et al., 2026i), and the AgileLens catalogue (Xu et al., 2026).

Of the 12 convincing systems, 11 are present in the primary Euclid Q1 strong lens catalogue, including the complete Einstein ring around NGC 6505 (O'Riordan et al., 2025). The remaining system, −573313189489592801 is a previously known strong lens reported by Euclid Collaboration et al. (2026c), having first been discovered in the DES Bright Arcs Survey (Diehl et al.,

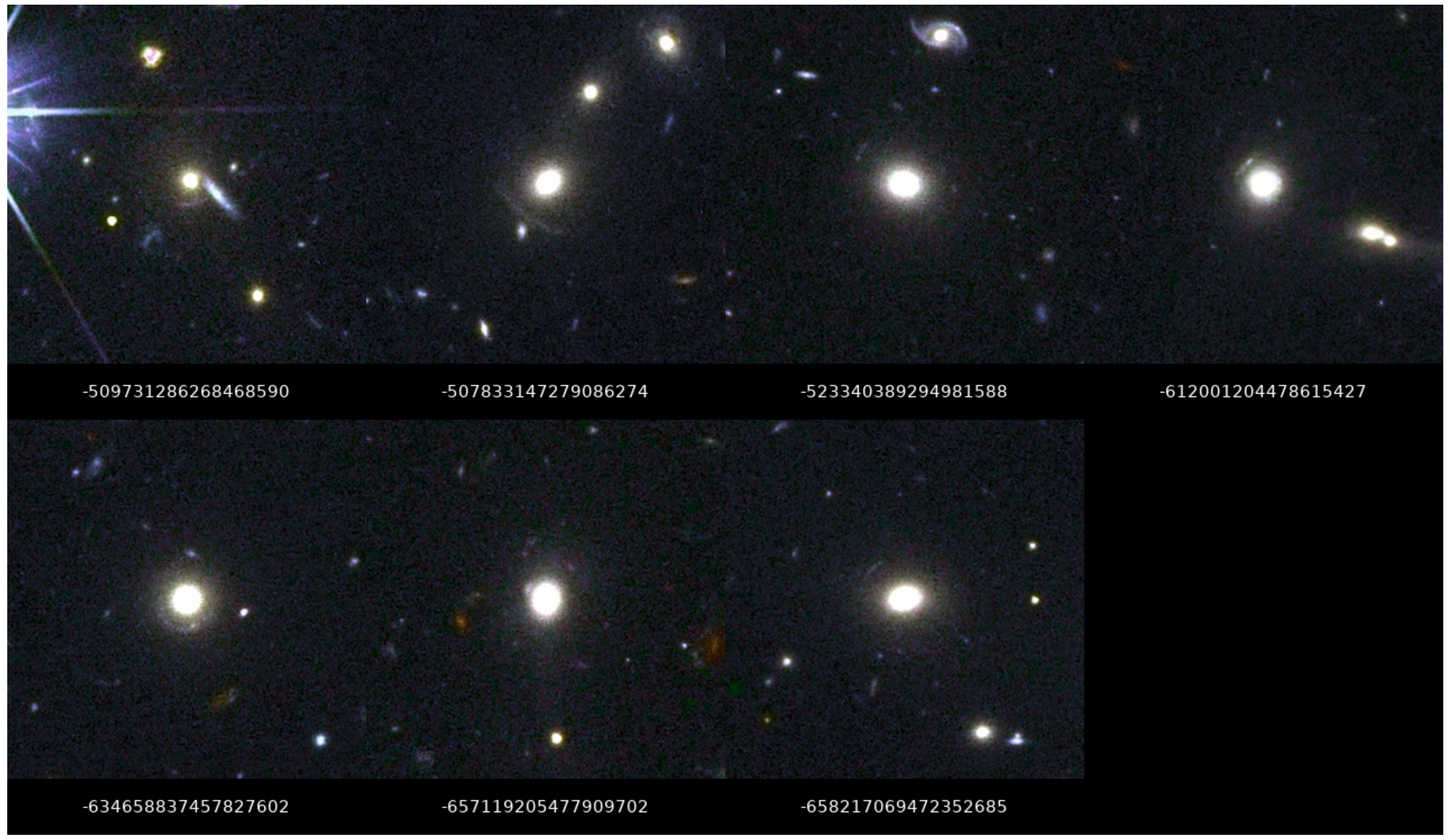

Figure 11: The seven visually selected strong lens systems with no counterparts in the published Euclid Q1 strong lens catalogues searched. All seven were retained as candidate systems following detailed visual inspection. Euclid object IDs are shown in each panel.

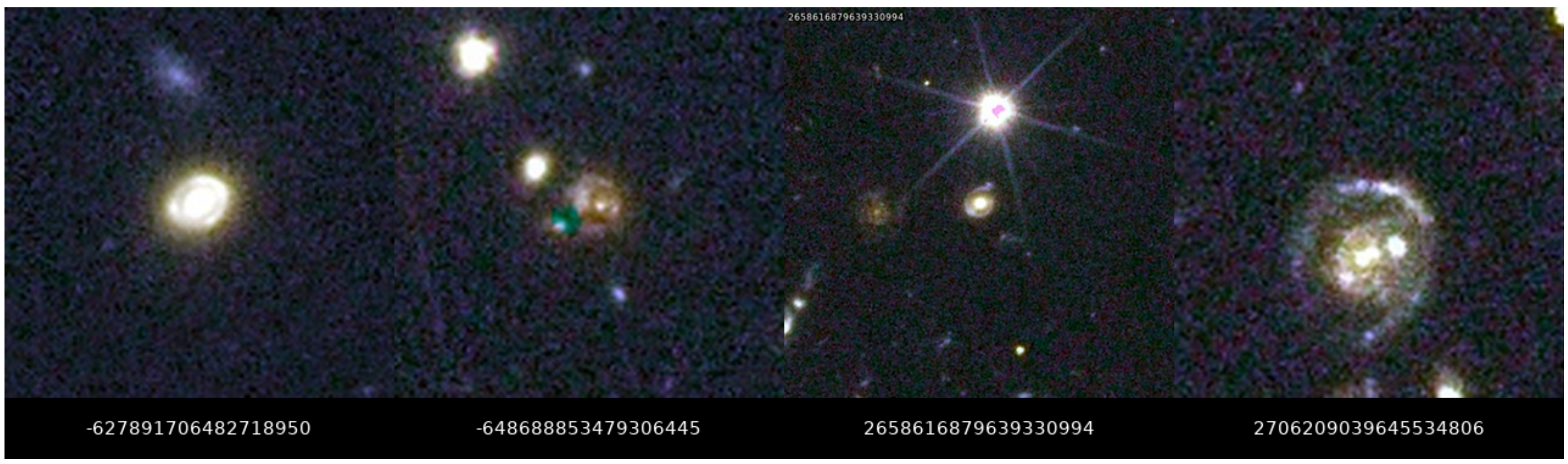

Figure 12: The four objects initially classified as ring-galaxy candidates by SS and subsequently reclassified as likely strong lenses following detailed visual inspection by SS and RN. Two of these systems are present in the primary Euclid Q1 strong lens catalogue, while no counterparts for the remaining two were found in the additional published Euclid Q1 strong lens catalogues searched. Euclid object IDs are shown in each panel.

2017).

Of the 17 candidate lenses, 8 are present in the primary Euclid Q1 strong lens catalogue. One of these, Euclid object ID 2704698077659154166 (EUCL J180152.75+655455.5), was also investigated by Euclid Collaboration et al. 2025. Subsequent lens modelling and redshift information in that work favour a superposition rather than a genuine strong lensing system. The remaining 9 candidate lenses exhibit less convincing or lower surface brightness lens like features and were retained as candidates.

In total, 9 of the retained systems have no counterparts in any of the published Euclid Q1 strong lens catalogues searched. Seven were identified among the objects initially classified as potential strong lenses and are shown in Figure 11, while the remaining two were initially classified as ring galaxies and are shown in Figure 12.

Of the 10 rejected systems, 5 were reclassified as ring galaxies (Figure 14) and transferred to the ring galaxy sample. The remaining 5 were rejected because the lensing feature was associated with a neighbouring source (1) or was attributed to pro-

-561927702490945297 -580346679490732536 -585825597474840222 -586076447512586764

-601410089495699776 -621119125483863792 -622410546466474464 -624735083485127883

-626405261479249819 2694329605673345769 2712125937662974309 2742362866663344665

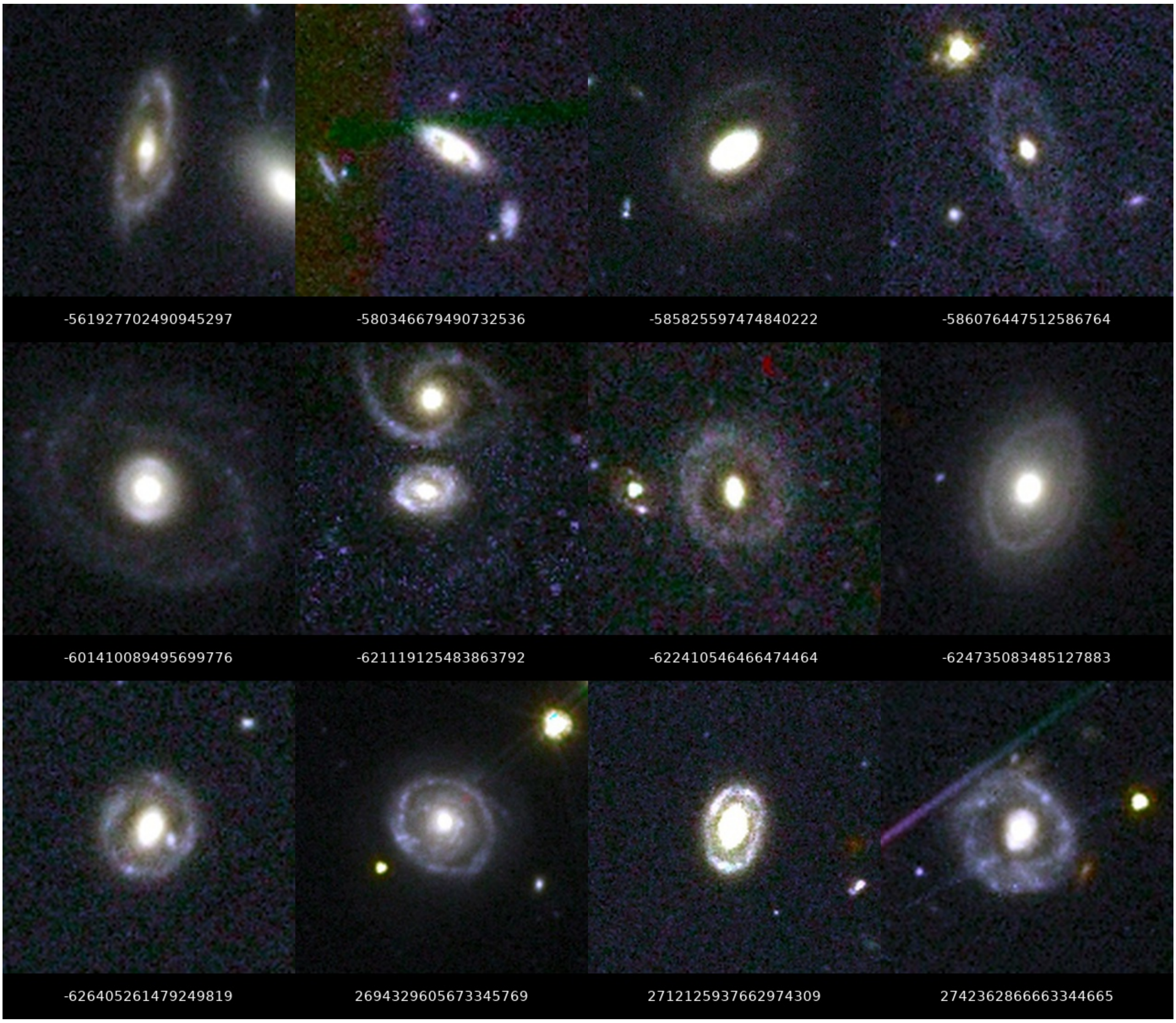

Figure 13: Representative examples of ring galaxies identified through the anomaly detection framework. The sample includes both classical, asymmetric and edge-on ring systems that provide a substantial sample for future studies of ring formation mechanisms, galaxy interactions, and triggered star formation.

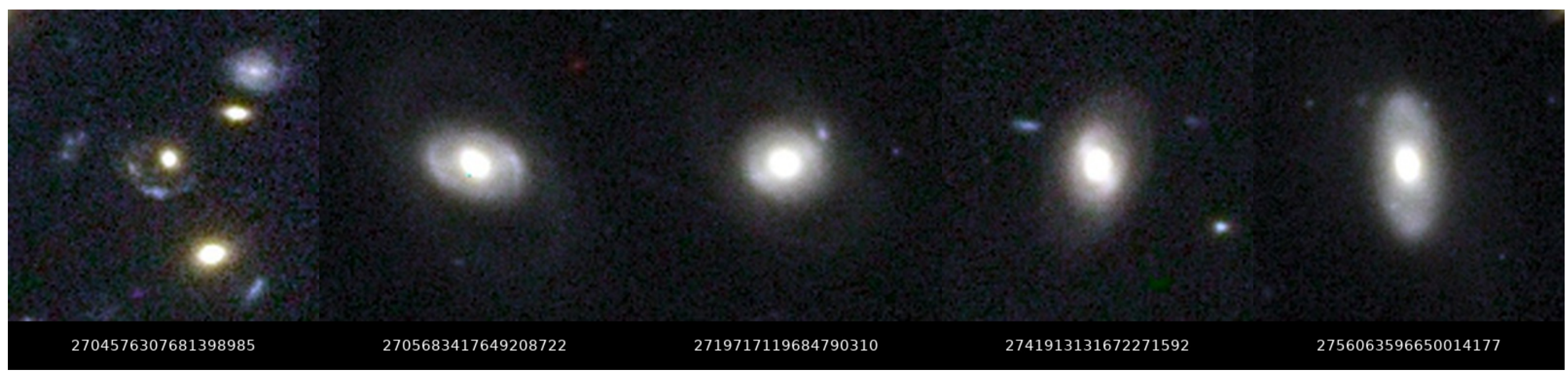

Figure 14: The five objects initially classified as potential strong lenses and subsequently reclassified as ring galaxies following detailed visual inspection. Their ring-like morphologies illustrate the visual similarity between ring galaxies and strong lensing systems, particularly those exhibiting near-complete Einstein rings. Euclid object IDs are shown in each panel.

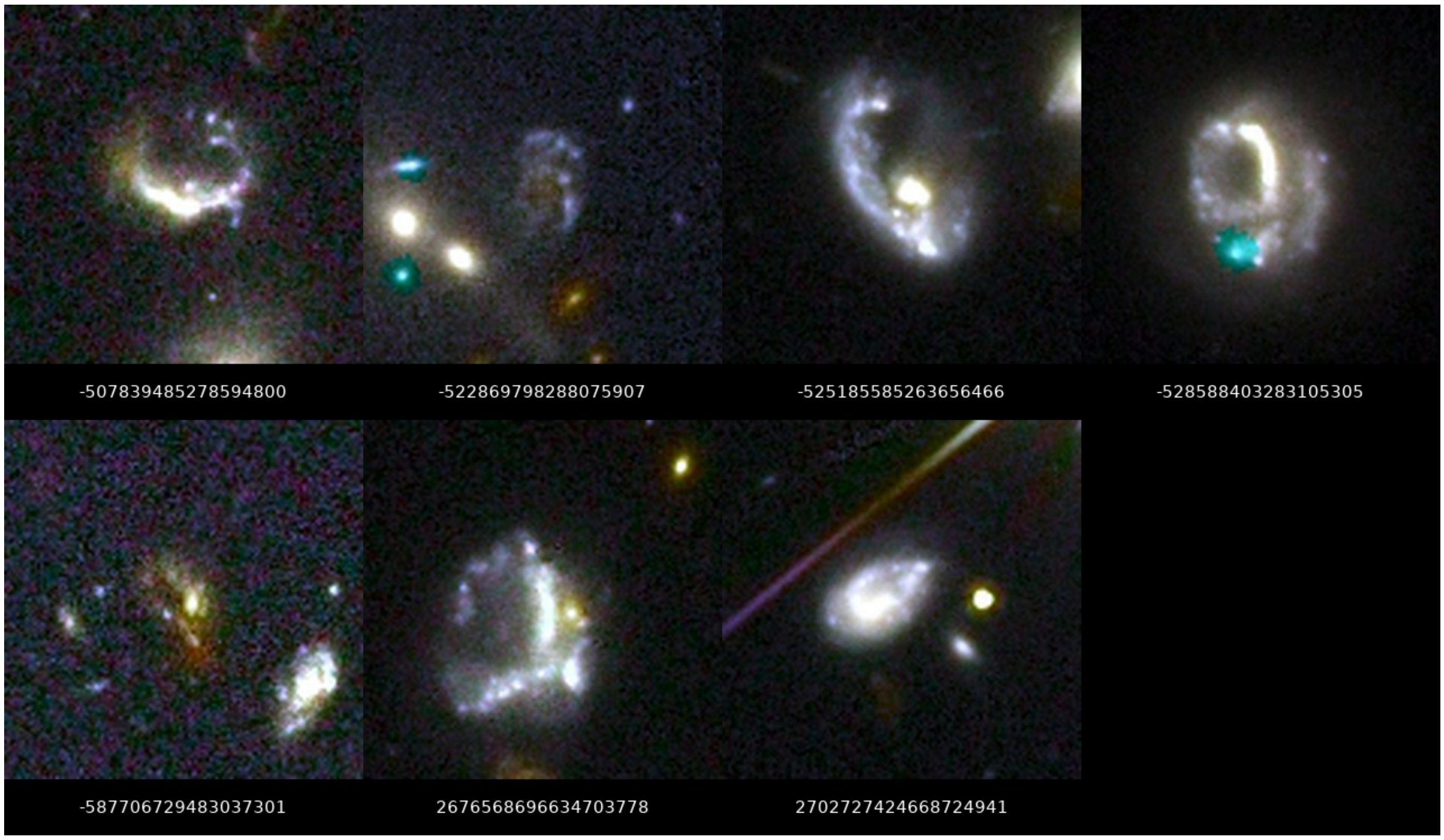

Figure 15: Representative examples of candidate collisional ring galaxies. These systems exhibit asymmetric ring-like morphologies and prominent star-forming knots suggestive of off-centre galaxy collisions. Some objects also display diffuse emission reminiscent of unconfirmed Voorwerp-like ionised structures.

jection or superposition effects (4). Representative examples of the visually convincing strong lens systems and candidate lenses are shown in Figure 10.

**Ring galaxies:** Of the 799 morphologically unusual objects recovered by the hybrid search, 214 were classified as ring-galaxy candidates during visual inspection, making this the largest morphological category in the recovered sample. Ring galaxies are characterised by prominent ring-like structures in their stellar discs, which may arise through a variety of dynamical processes (Buta and Combes, 1996). Here, the classification is purely morphological and does not imply a specific formation mechanism.

Following detailed visual inspection by SS and RN, 197 of the initial candidates were retained as convincing ring galaxies. Four objects were instead reclassified as likely strong lenses. Two of these (Euclid object IDs 2658616879639330994 and −627891706482718950) are present in the primary Euclid Q1 strong lens catalogue (Euclid Collaboration et al., 2026j), while no counterparts for the remaining two were found in the additional published Euclid Q1 strong lens catalogues searched. The four reclassified systems are shown in Figure 12.

Including the 5 systems reclassified from the strong lens sample (Figure 14) the final sample comprises 202 visually classified ring galaxies. The remaining 13 initial ring candidates were rejected as either class, with their apparent ring like structures instead attributable to features such as faint spiral arms. Representative examples are shown in Figure 13.

Larger ring-galaxy catalogues exist in the literature, including 3962 ringed galaxies identified from Galaxy Zoo 2 (Buta, 2017) and 33,993 ring-galaxy candidates identified from HSC imaging using a deep-learning classifier (Shimakawa et al., 2024). In contrast, we are not aware of a dedicated ring-galaxy catalogue from Euclid Q1. Ring galaxies identified in the Q1 strong lensing analyses have primarily been treated as non lenses or false positives. The present sample therefore provides an independently selected set of ring morphologies in Euclid imaging that can be used for subsequent morphological and population studies.

**Collisional ring galaxies:** Twelve objects were initially assigned to the category of collisional ring galaxies by SS (Appleton and Struck-Marcell 1996; O'Ryan and Gómez 2025), characterised by asymmetric ring-like structures and prominent star-forming knots suggestive of off-centre galaxy collisions. Following detailed visual inspection by SS and RN, seven objects were retained in this class, while the remaining five were reclassified as mergers, peculiar galaxies, or disturbed spirals. The seven retained objects are shown in Figure 15.

Collisional ring galaxies are intrinsically rare, with existing samples typically containing only a few tens to approximately one hundred systems. The catalogue of Madore et al. (2009), for example, contains 127 collisional ring systems, while searches at higher redshift have identified samples of 25 (Lavery et al., 2004) and 24 (Elmegreen and Elmegreen, 2006) candidate ring systems. The 7 systems recovered here therefore constitute a non-negligible sample given the limited area covered by Euclid Q1. Several candidates also exhibit diffuse emission reminiscent of the ionised clouds associated with Voorwerp-like systems (Lintott et al. 2009; Keel et al. 2012). Broadband imaging alone cannot distinguish between stellar light associated with disturbed collisional rings and emission line gas photoionised by a fading AGN, and we therefore retain these systems as morphological candidates.

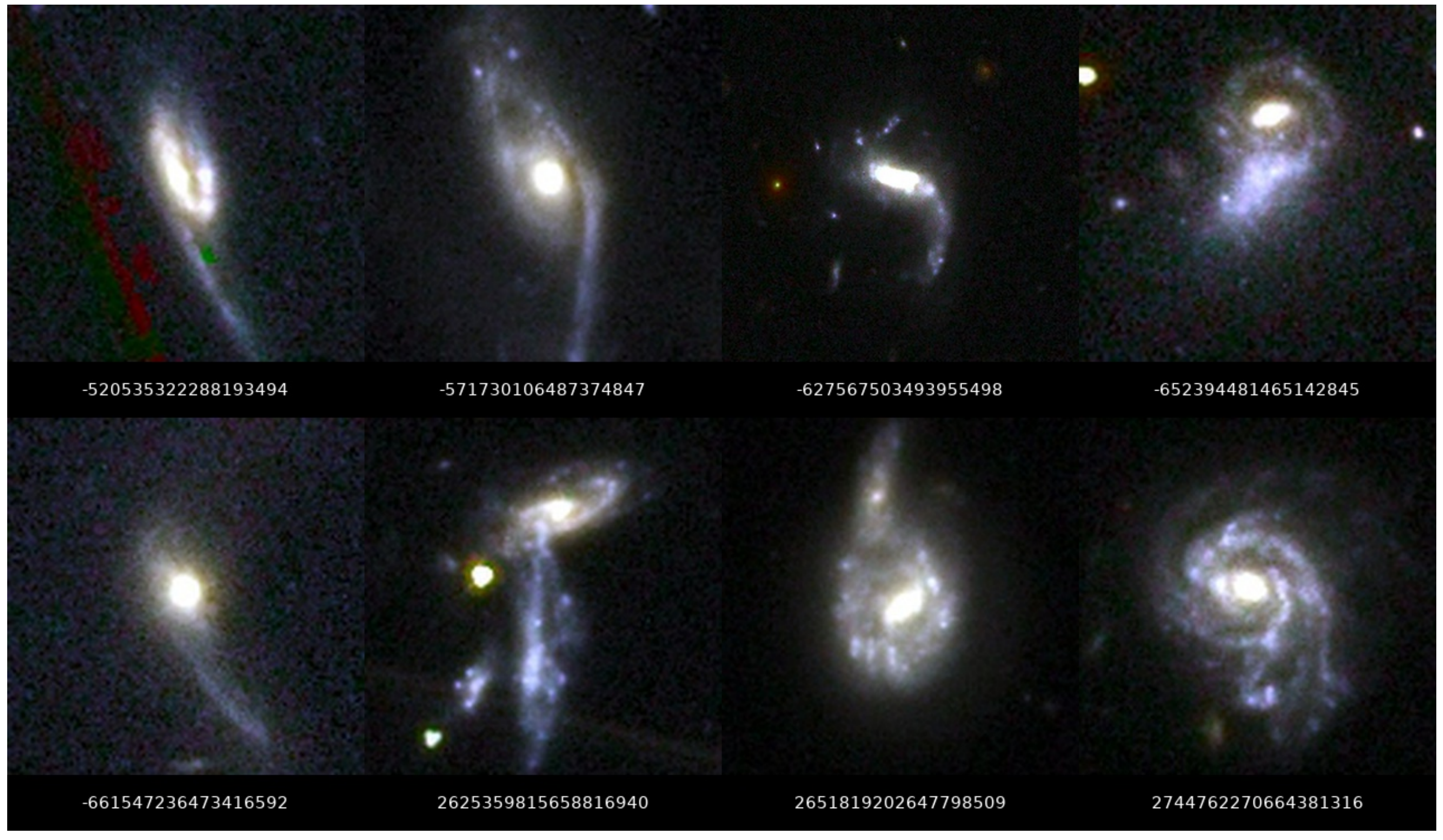

Figure 16: Representative examples of jellyfish-like one sided asymmetric galaxy candidates. These galaxies exhibited pronounced morphological asymmetry, frequently characterised by diffuse emission or extended stellar structure predominantly on one side of the galaxy and can be regarded as candidate systems pending spectroscopic and environmental follow-up.

**Jellyfish-like one-sided asymmetric galaxies:** Jellyfish galaxies are systems undergoing gas stripping as they move through a dense intracluster or intragroup medium, most commonly as a result of ram pressure stripping. The interaction between the interstellar medium of the galaxy and the surrounding hot gas can remove material from the disc, producing characteristic one sided tails of gas and regions of enhanced star formation that give these systems their "jellyfish" appearance (Gunn and Gott III 1972; Poggianti et al. 2017).

Here, the term *jellyfish-like* is used as a morphological description rather than a physical classification. This category was defined broadly to include galaxies exhibiting pronounced one-sided morphological asymmetry, often characterised by diffuse stellar or gaseous emission apparently extending from one side of the galaxy, giving the visual impression of a jellyfish galaxy (Poggianti et al. 2017; Nardone et al. in prep.). Representative examples are shown in Figure 16.

A total of 66 objects were initially assigned to this category (SS). Following detailed visual inspection (SS and RN), 52 were retained as one-sided asymmetric galaxy candidates, one object was considered uncertain, and 13 were rejected because their morphologies were more consistent with tidal interactions, mergers, or asymmetric spiral structure.

For comparison, the systematic low-redshift search of Poggianti et al. (2016) identified 344 stripping candidates in 71 clusters and a further 75 in groups and lower-mass environments, while searches at higher redshift have typically produced samples of order one hundred objects. The 52 systems recovered here therefore represent a substantial sample given the limited area covered by Euclid Q1. A simple area scaling from Q1 ($63.1, \mathrm{deg}^2$) to the expected $\sim 1900, \mathrm{deg}^2$ coverage of DR1 would correspond to approximately 1500–1600 jellyfish-like candidates, assuming a comparable source density and selection function (Euclid Collaboration et al., 2026b). **Galaxy interactions and their remnants:** Galaxy interactions and mergers play an important role in galaxy evolution, with the potential to alter galaxy morphology, redistribute gas and stars, modify star formation, and contribute to nuclear activity (Toomre and Toomre 1972; Barnes and Hernquist 1996; Conselice 2014). The merger rate and its dependence on galaxy properties and environment therefore provide important constraints on models of galaxy formation and evolution (Robotham et al., 2014). Observable signatures of an interaction also evolve throughout the encounter, from close pairs with distinct components to strongly disturbed coalescing systems and, eventually, single remnants that retain tidal signatures of the interaction. We therefore divided the systems recovered here into three broad morphological categories representing different stages of this sequence.

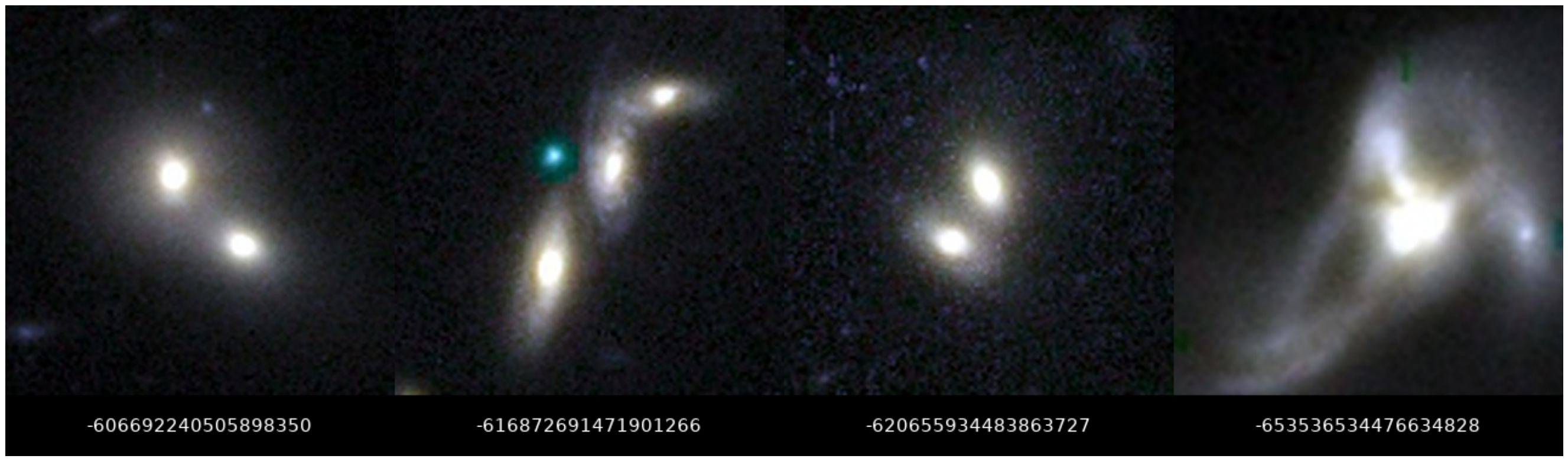

(a) Merging or close galaxies

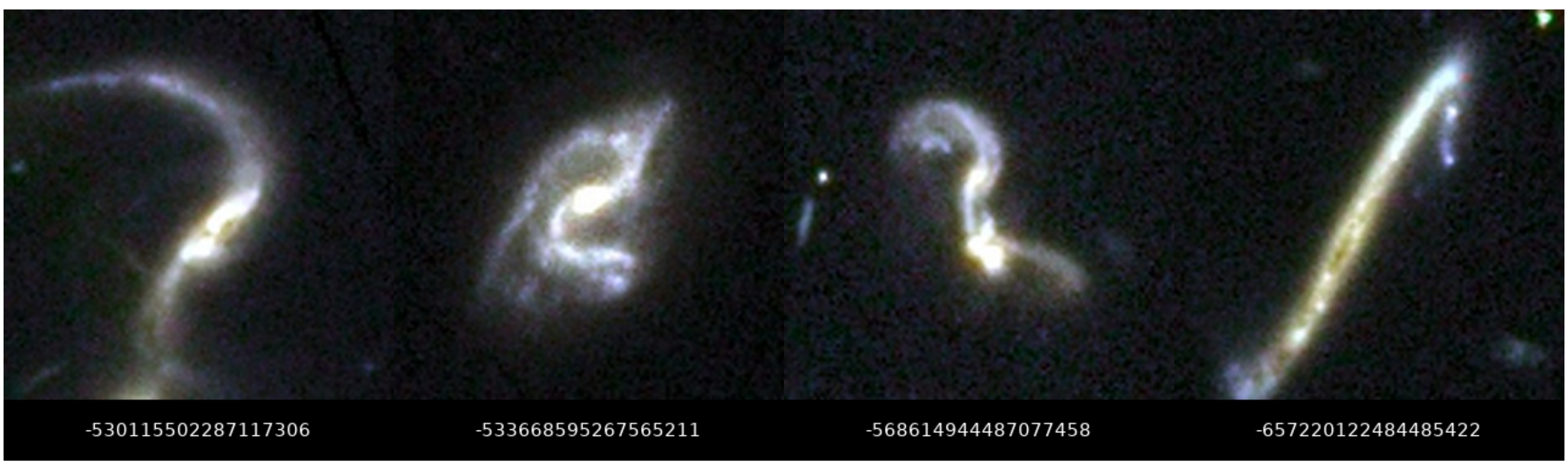

(b) Advanced mergers

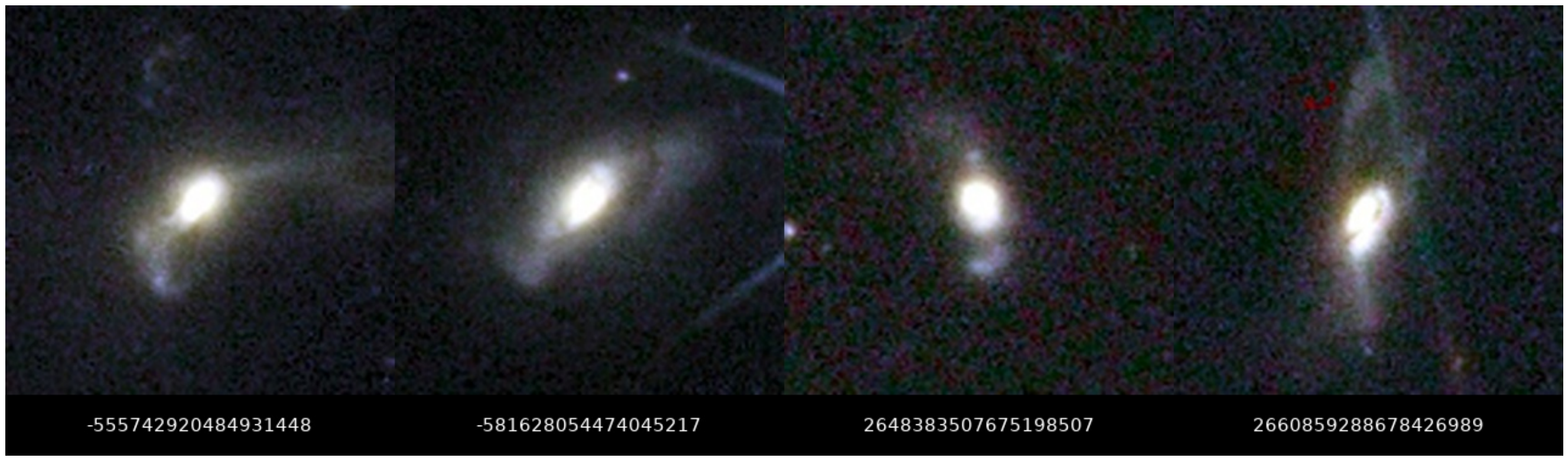

(c) Merger remnants

Figure 17: Representative examples of interacting galaxies recovered by the anomaly detection framework. (a) Merging or close galaxies. (b) Advanced mergers. (c) Merger remnants. Together, these categories span the interaction sequence from the earliest stages of gravitational interaction through coalescence to the disturbed remnants left by major mergers.

*Merging or close galaxies.* This category comprises galaxy pairs in which the individual galaxies remain morphologically distinct but exhibit evidence consistent with gravitational interaction. Typical signatures include small projected separations, tidal bridges, distorted outer isophotes, asymmetric spiral arms, and tidal tails, while the individual components remain separately identifiable. A total of 160 objects were initially assigned to this category, and all 160 were retained following detailed visual inspection.

*Advanced mergers.* This category contains systems in which the progenitor galaxies appear to have substantially coalesced, although prominent signatures of the interaction remain visible. These include highly disturbed stellar morphologies, tidal tails, shells, loops, warped structures, and, in some cases, multiple or poorly separated central components. A total of 24 objects were initially assigned to this category, and all 24 were retained

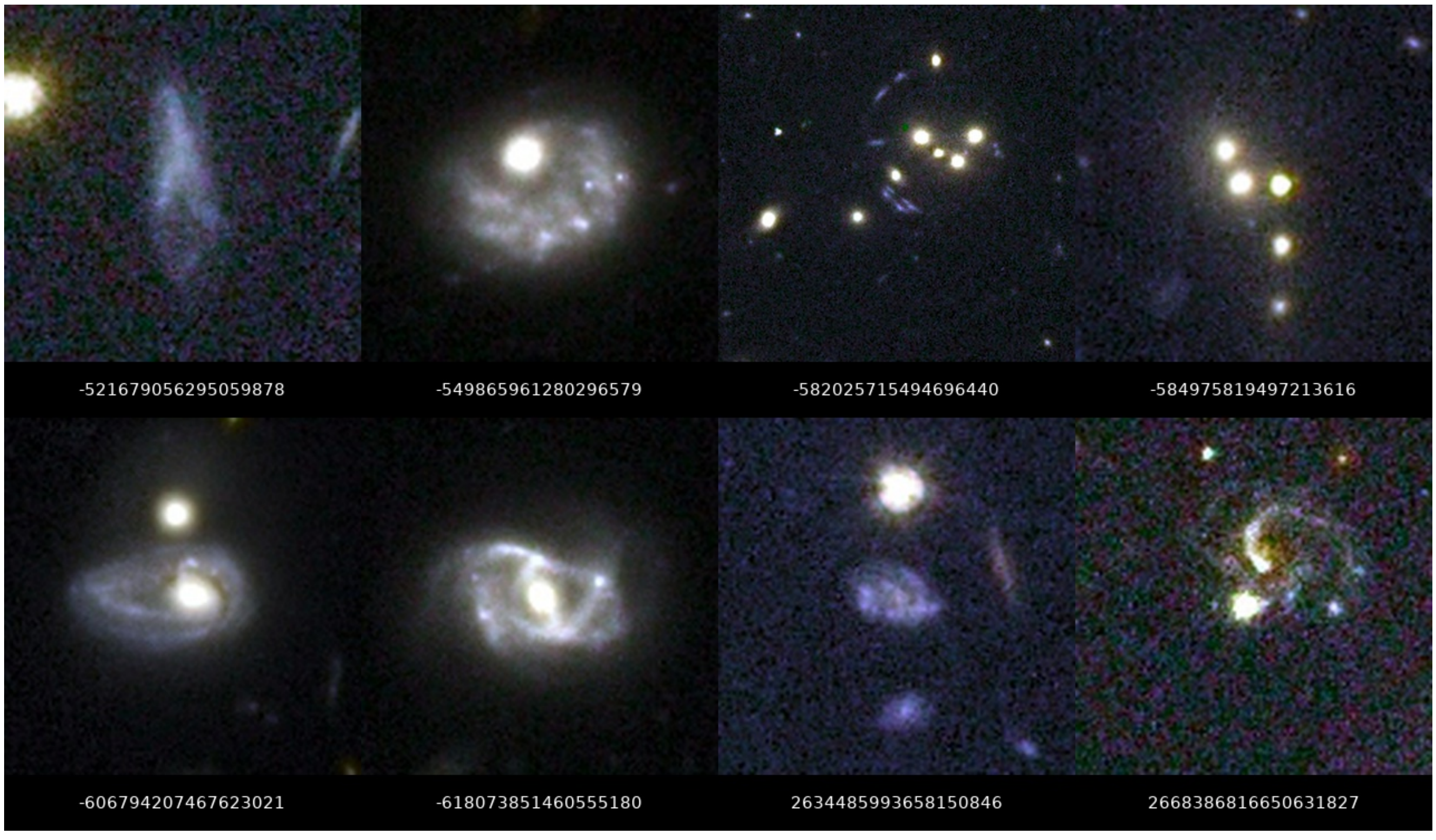

Figure 18: Representative examples of morphologically peculiar galaxies identified by the anomaly detection framework. The objects exhibit a wide variety of unusual morphologies, including irregular stellar structures, compact multiple systems, unusual ring-like morphologies, and complex interaction geometries, highlighting the ability of the framework to recover rare objects that fall outside conventional morphological classes.

following detailed visual inspection.

*Merger remnants.* This category contains systems dominated by a single remnant while retaining morphological signatures consistent with a recent interaction. Common features include tidal tails, shells, plumes, diffuse stellar envelopes, asymmetric outer isophotes and, in some cases, residual double nuclei. Such faint tidal structures can preserve evidence of previous interactions after the main bodies of the progenitor galaxies have coalesced (Walmsley et al., 2019). A total of 86 objects were initially assigned to this category, and all 86 were retained following detailed visual inspection.

Together, the three categories contain 270 systems, examples of which are shown in Figure 17.

**Morphologically peculiar systems:** The final category contains morphologically peculiar systems that could not be confidently assigned to any of the preceding classes. These objects exhibit a diverse range of uncommon structures, including highly irregular morphologies, compact multiple systems, unusual ring-like features, diffuse stellar envelopes, and complex interaction geometries. Their heterogeneous nature precludes a single physical interpretation, but collectively they represent some of the most unusual discoveries produced by the anomaly detection framework. A total of 202 objects were assigned to this category, and retained following visual inspection. Representative examples are shown in Figure 18.

## 6. Discussion and Conclusions

Conventional anomaly detection ranks objects according to statistical unusualness without knowledge of their scientific relevance. In A-AD, the definition of an interesting anomaly is progressively shaped by expert feedback (Lochner and Bassett, 2021; Kornilov et al., 2025). By applying Astronomaly, Protege, and PineForest to identical feature representations within each dataset, we show that the way this feedback is incorporated also matters. On D1, Protege consistently achieved the highest anomaly recovery, followed by PineForest and Astronomaly, whereas on D2, PineForest prioritised the adopted anomaly definition faster than the other two. The effect of prior information was itself algorithm dependent: recovery varied with the composition of the supplied prior sample, although these variations were generally smaller than those produced by the choice of A-AD algorithm. Thus, even when operating on the same representation of the same astronomical sources, different A-AD strategies can produce different recovery efficiencies and discovery trajectories. Benchmarking an A-AD method on a single dataset or modality may therefore provide an incomplete assessment of its performance. The Euclid experiment illustrates how such a discovery search differs from a targeted classification problem. Strong gravitational lenses provide a particularly clear example. Dedicated lens searches are explicitly designed or trained to identify lensing systems and can be evaluated in terms of their completeness and purity (Euclid Col-laboration et al. 2026f,e). In contrast, neither the initial anomaly ranking nor the feature representation used here contained an explicit definition of gravitational lensing. Nevertheless, the hybrid search recovered known strong lenses together with candidates without counterparts in the Euclid Q1 lens catalogues searched in this work (Section 5.2). It also recovered ring galaxies, collisional-ring candidates, jellyfish-like systems, mergers, and other peculiar morphologies, none of which were predefined target classes of the machine-learning algorithms. A-AD therefore provides a complementary route to targeted searches by allowing unusual systems to emerge without requiring every scientifically interesting class to be specified in advance.

This flexibility also determines how the resulting samples should be interpreted. The experiments were not designed to establish formal selection functions, and the human labels necessarily encode the scientific judgement of the classifiers. Moreover, the visual classifications are not definitive physical classifications: similar morphologies can arise from different phenomena, as illustrated by the overlap between ring galaxies and Einstein rings and between jellyfish-like morphologies and other asymmetric galaxies. The resulting samples should therefore be regarded as morphologically selected candidate samples rather than complete or unbiased catalogues. At the same time, the algorithm-dependent behaviour can be used constructively. In the hybrid experiment (Section 5), PROTEGE was used to interactively construct a labelled sample that was subsequently transferred to PINEFOREST as prior information for large scale discovery, combining the interactive recommendation strategy of PROTEGE with the computational scalability of PINEFOREST. To our knowledge, this is the first demonstration in astronomy of expert knowledge accumulated through one A-AD framework being transferred to another to drive a large-scale discovery search.

The experiments also produced candidate discoveries across both data modalities. In D1, among the first 200 objects inspected for each method in the Baseline Comparison, PINEFOREST, ASTRONOMALY and PROTEGE recovered 39, 12 and 15 previously uncatalogued variable star candidates respectively. In D2, the hybrid PROTEGE–PINEFOREST search identified 799 morphologically unusual objects among the 10,000 Euclid images inspected. Following detailed visual classification, these included 29 retained strong lens systems and candidates, comprising 12 visually convincing systems and 17 candidates, as well as 202 ring galaxies, 7 collisional ring galaxies, 52 jellyfish-like systems, 270 merging or interacting systems, and 202 other morphologically peculiar galaxies. Of the 17 strong lens candidates, 9 have no counterparts in the published Euclid Q1 strong lens catalogues searched. Seven were identified directly among the potential strong lenses, while 2 were initially classified as ring galaxies and subsequently reclassified as strong lens candidates. We are also not aware of a dedicated Euclid Q1 catalogue for the 202 visually classified ring galaxies.

Two main conclusions emerge from this work. First, A-AD performance is not algorithm independent: different mechanisms for incorporating expert feedback can lead to different scientific discoveries even when the input data and feature representation are held fixed, while the effectiveness of prior information also depends on its composition and on the A-AD method to which it is supplied. Second, these differences can be used constructively by transferring expert knowledge between complementary A-AD methods, providing a route from interactive exploration to large-scale discovery. For surveys such as Euclid and the Vera C. Rubin Observatory LSST, where unexpected phenomena cannot always be specified in advance, this offers a complementary approach to both unsupervised anomaly detection and targeted classification.

## Acknowledgements

We thank Dr. Maria V. Pruzhinskaya for valuable discussions on the light-curve and image data, on the design and evaluation of the experiments and the paper structure. We also thank Dr. Margherita Grespan for helpful discussions regarding the Euclid image data.
This work has made use of the Quick Release (Q1) data from the Euclid mission of the European Space Agency (ESA). The Q1 dataset was processed by the Euclid Consortium.